\documentclass[12pt]{article}
\pdfoutput=1
\usepackage{jheppub}

\usepackage{bm}
\usepackage{ifsym}
\usepackage{amsthm}
\usepackage{amsfonts}
\usepackage{ccaption}
\usepackage{mathrsfs}
\usepackage{booktabs}

\makeatletter
\@addtoreset{equation}{section}

\makeatletter
\renewcommand\section{\@startsection {section}{1}{\z@}%
                                   {-3.5ex \@plus -1ex \@minus -.2ex}%nn
                                   {2.3ex \@plus.2ex}%
                                   {\normalfont\large\bfseries}}
\renewcommand\subsection{\@startsection{subsection}{2}{\z@}%
                                     {-3.25ex\@plus -1ex \@minus -.2ex}%
                                     {1.5ex \@plus .2ex}%
                                     {\normalfont\bfseries}}

  \captionnamefont{\bfseries}
  \captiontitlefont{\small\sffamily}
  \captiondelim{: }
  \hangcaption

\def\sec#1{\S\ref{#1}}
\def\fig#1{Fig.\,\ref{#1}}
\def\req#1{(\ref{#1})}
\def\App#1{Appendix \ref{#1}}

\definecolor{maroon}{rgb}{0.8,0.2,0.2}
\definecolor{grey}{rgb}{0.7,0.6,0.7}

\def\mik{outgoing} % Mikhailov's boundary condition for WS
\def\aq{anti-quark}

\def\tg{{t}}  
\def\ph{\varphi}

\def\dsWS{ds_{\gamma}}  % worldsheet metric
\def\scri{\mathscr I}

\def\rmin{r_{m}}
\def\phq{\ph_q}
\def\phaq{\ph_{\tilde q}}

\def\RR{\mathbb{R}}

\title{String worldsheet for accelerating quark}

\author{Veronika E. Hubeny$^a$}
\author{\& Gordon W. Semenoff$^b$}

\affiliation[a]{ Centre for Particle Theory \& Department of Mathematical Sciences,\\
Science Laboratories, South Road, Durham DH1 3LE, UK.}
\affiliation[b]{Department of Physics and Astronomy, University of British Columbia, \\
                     6224 Agricultural Road, Vancouver, British Columbia, Canada V6T 1Z1}
\emailAdd{veronika.hubeny@durham.ac.uk}
\emailAdd{gordonws@phas.ubc.ca}

\abstract{
We consider the AdS bulk dual to an external massive quark in SYM following an arbitrary trajectory on Minkowski background.  While a purely outgoing boundary condition on the gluonic field allows one to express the corresponding string worldsheet in a closed form, the setup has curious consequences. In particular, we argue that 
any quark whose trajectory on flat spacetime approaches that of a light ray in the remote past (as happens e.g.\ in the case of uniform acceleration)
must  necessarily be accompanied by an \aq.
This is puzzling from the field theory standpoint, since one would expect that a sole quark following any timelike trajectory should be allowed.  We explain the resolution in terms of boundary and initial conditions.
We analyze the configuration in global AdS, which naturally suggests a modification to the boundary conditions allowing for a single accelerated quark without accompanying \aq.  We contrast this resolution with earlier proposals.
 } 

\begin{document}
\begin{flushright} \small{DCPT-14/43} \end{flushright}

\maketitle

\flushbottom
%\renewcommand{\thefootnote}{\arabic{footnote}}
%______________________________________

%____________________________________________
\section{Introduction}
\label{s:intro}
%____________________________________________

Among the many merits of AdS/CFT correspondence\footnote{
For definiteness we'll consider the prototypical case of  AdS/CFT correspondence \cite{Maldacena:1997re} 
which relates the 4-dimensional ${\cal N}=4$ Super Yang-Mills  (SYM) gauge theory to a IIB string theory on asymptotically AdS$_5 \times S^5$ spacetime; in our considerations below, the $S^5$ will not enter the story in any nontrivial way.
} is the geometric insight it provides into strongly coupled field theory.
For example, it allows one to analyze the strong coupling planar limit of a gauge theory by studying the classical gravitational physics of its dual.   An application which has received considerable attention
is  the dynamics of a heavy quark which is inserted into ${\cal N}=4$ supersymmetric Yang Mills theory, particularly
its behavior in real-time and out-of-equilibrium situations where direct analytical or numerical tools in the context
of conventional gauge theory are still lacking.  
In the gauge theory we might imagine specifying the trajectory 
of the quark and analyzing, for example,
 the power that is required to force the quark to follow its worldline,
how much radiation it produces in response, how this radiation propagates, etc..\footnote{
See e.g.\ \cite{CasalderreySolana:2011us,Adams:2012th} for reviews of this AdS/QCD programme.}

In the gravity dual, the quark is the endpoint of an open string.  For simplicity we will consider the quark propagating through vacuum state of the SYM plasma, corresponding to pure AdS geometry in the bulk.  The quark itself lives at the
asymptotic boundary\footnote{
Here for simplicity we assume that we have an infinitely massive quark (corresponding to a bare source in the fundamental representation of the gauge group);  a finite-mass quark could be easily accommodated by introducing a probe D-brane into AdS$_5$ which caps off at correspondingly large radius, effectively regulating the string length.  However, since the portion of the string `below' the brane is identical in both cases, it will suffice to consider the simpler infinite mass setup. 
}  of AdS while the string hangs into the bulk,
and encodes both the gluonic cloud of the quark as well as its radiation field.
The classical geometry of the worldsheet is
determined by extremizing its proper area subject to the boundary condition that it ends on the quark worldline.  
However, this specification is incomplete without further input.  
A fundamental open string has two endpoints.  In global AdS, both endpoints lie on the boundary; due to the string's orientation, the second endpoint codifies an anti-quark.
One natural way to complete the specification of the string worldsheet, then, is to impose Dirichlet boundary conditions for both ends of the worldsheet, which amounts to specifying worldlines of both the quark and \aq.   This makes the problem for finding the extremal worldsheet well posed (albeit computationally nontrivial). 

However, we might be interested in just a single quark and its gluonic field without explicitly specifying the behavior of the \aq.  There is a very natural boundary condition, analogous to the familiar one used in determining the electric field of a moving charge in electrodynamics, which ensures that the quark's behavior at a given time influences the worldsheet only along its future lightcone.  
In particular, any excitation sourced by the quark is precisely sinked by the \aq\ at the other end, so that no excitation can propagate back  towards the quark.  Since in the field theory this corresponds to purely outgoing gluonic field configuration (as befits any radiation sourced by an accelerating charge), we will refer to this boundary condition as `\mik'.\footnote{
Note that, consistently with the naive UV/IR intuition, the corresponding string excitation  is then purely {\it ingoing} in AdS, i.e.\ propagating along the string into the bulk, away from the boundary.
}

The \mik\ boundary condition likewise provides a complete specification of the problem, allowing us to determine the string worldsheet (in the causal future of the quark's specified trajectory).  Physically it implements the expectation that as the endpoint is accelerated, the rest of the string causally trails behind as it hangs into the bulk.
This boundary condition was crucially utilized by Mikhailov \cite{Mikhailov:2003er} to determine the exact solution for string worldsheet in Poincare-AdS in terms of the quark trajectory on the boundary Minkowski spacetime.  Indeed, since the worldsheet is ruled by null geodesics, the solution can be compactly written in a closed form for {\it arbitrary} quark trajectory, as we review in \sec{s:setup}.  

In global AdS$_5$, whose boundary is the Einstein Static Universe (ESU) $S^3 \times R^1$, we see the entire string; the field theory lives on a compact space, so that it must satisfy the Gauss law constraint:  it cannot admit a quark without an \aq.  However, if we restrict attention to just a single Poincare patch of AdS, whose boundary is the spatially non-compact Minkowski space, the string could extend through the Poincare horizon, rendering only one endpoint visible on the Minkowski boundary corresponding to this patch.  In such a case, it should be possible to consider a single quark in isolation.  
After all, the conformally symmetric Yang-Mills theory is in a deconfined phase, so individual quarks should be allowed, and one might expect that the absence of Gauss law constraint on the non-compact space should indeed allow color charged states, including a quark which follows {\it any} timelike trajectory, without the presence of \aq.

Nevertheless, a naive implementation of the \mik\ boundary conditions seems to contradict this expectation.  In particular, for sufficient quark acceleration, the corresponding string could whip around and actually reemerge on the boundary within the same Poincare patch.  The simplest and best-known example of this phenomenon occurs for a uniformly-accelerating quark (as has been studied many times in the past starting with \cite{Xiao:2008nr}),
which we review in \sec{s:unifaccel} below.  
However, this occurrence is far more general: as we explain in \sec{s:aqcriterion}, an \aq\ is induced whenever the quark undergoes past-eternal acceleration so as to approach a light ray.\footnote{
Although perhaps not entirely obvious in the context of the field theory on flat spacetime, such trajectories are in fact {\it generic}, as we will become more evident in the global viewpoint.  }  Hence for a rather generic class of quark trajectories (for which the acceleration does not vanish too quickly at early time), it seems that {\it we cannot consider a single quark on Minkowski spacetime, because the bulk string creates a `mirror' \aq.}
Its existence, which is apparently required by the string theory, then presents a puzzle on the Yang-Mills side of the duality.

Clearly the subtlety lies with the boundary/initial conditions.  This is perhaps not so surprising, as conundrums involving eternal (as opposed to merely arbitrarily long) acceleration have long and rich history.\footnote{
In fact, the analogous question in electrodynamics has been studied for over a century; see e.g.\ \cite{bondi1955field} for review of early discussions regarding the field of single uniformly accelerated charge versus a pair of opposite uniformly accelerated charges.
}
One possible resolution has recently been explored by
\cite{Garcia:2012gw}, who propose that the correct continuation of the string worldsheet is non-smooth so as to encode a gluonic shock wave accompanied by the onset of quark's acceleration at `initial' time.  Part of the string worldsheet (which cannot be causally affected by the uniformly accelerating quark at any finite time in the past) is null, matching onto a timelike part of the worldsheet presented by \cite{Xiao:2008nr}.  
Here we offer a very different resolution to that proposed by \cite{Garcia:2012gw}, one which is naturally suggested by the global AdS viewpoint:  we can simply consider a different set of boundary conditions, such as the one mentioned above invoking Dirichlet conditions on both endpoints, and force the \aq\ end of the string to pass through the point of the global AdS boundary sphere which is invisible to the Poincare patch.  This allows for a smooth worldsheet; in fact for uniformly accelerating quark the entire string configuration is globally static.

The crux of our bulk analysis is actually very simple.
While the original setup (which considers a given quark trajectory on flat spacetime) is most naturally explored in Poincare AdS, it turns out that it is conceptually most convenient to view the problem in global AdS,  restricting attention to the Poincare patch only at the end of the calculation.  In this setting, we can keep track of both string endpoints, and formulate a simple criterion for obtaining a single quark  versus a quark-\aq\ pair.
 Apart from providing a perhaps more intuitive picture of how the string responds to the quark's motion, this visualization in global AdS also reveals interesting symmetries which are not manifest in the Poincare patch, such as a global rotation which relates a configuration of a single static quark to that of the uniformly accelerating quark-\aq\ pair. 

At the danger of being pedantic, it may be useful to emphasize the distinction between the various spacetimes that we will encounter.  In order of decreasing dimensionality, we will describe
\begin{itemize}
\item The five-dimensional {\it bulk} AdS spacetime.  We restrict attention to pure AdS (corresponding to the vacuum state of the YM), but we will consider both the full global AdS as well as its Poincare patch.
\item
The four-dimensional {\it boundary} of AdS, conformal to the spacetime on which the field theory `lives'.  In case of global AdS, we think of the field theory as living on ESU, while for Poincare AdS we think of the field theory as living on Minkowski background.
\item
The two-dimensional open string {\it worldsheet} inside AdS, which ends on the AdS boundary.
\end{itemize}
All three spacetimes are Lorentzian and causal.  Furthermore, the geometry of the first two is fixed; for our discussion, we don't need to consider the backreaction of the string on AdS.  The worldsheet spacetime geometry can vary, but this is rather trivial due to the low dimensionality.  The main content, then, will come from global considerations of the effects of the boundary conditions.

The plan of the paper is as follows.
In \sec{s:setup} we present the general set-up obtaining string worldsheet specified by an arbitrary quark trajectory with \mik\ boundary conditions.  We start by explicating the properties of the worldsheet geometry (such as the presence of worldsheet horizon)  in \sec{s:WS}.  Although most of these results are simple and appeared in the literature before, we include them for completeness and pedagogical value.  We then show how to view the string in global AdS in \sec{s:globalAdS}, and use this to formulate a criterion on the quark trajectory for which a smooth extension of the string worldsheet with \mik\ boundary conditions induces an \aq\ in \sec{s:aqcriterion}.
To illustrate this criterion, in \sec{s:unifaccel} we focus on the most familiar example of uniform acceleration.
In \sec{s:resolution} we discuss how to contrive a scenario admitting a uniformly accelerating quark without accompanying \aq; we present our proposal in \sec{s:bentWS} and then compare it with the previous discussions in \sec{s:resolcomp}.  We end with closing remarks in 
\sec{s:discussion}.
To keep mathematical clutter to a minimum in the main text, we relegate some of the explicit expressions and generalizations (as well as corresponding figures) to \App{s:arbacc}.

%____________________________________________
\section{String worldsheet for arbitrary quark worldline}
\label{s:setup}
%____________________________________________

Let us start by considering a quark moving on Minkowski spacetime $ds_{bdy}^2 = \eta_{\mu \nu} \, dx^\mu \, dx^\nu$.  One can specify the quark's worldline by $x^\mu(\tau)$, parameterized by  proper time $\tau$.  This worldline can be arbitrary provided it is timelike everywhere, and the parameterization by $\tau$ ensures that the four-velocity $\dot{x}^\mu(\tau) \equiv \frac{d x^\mu(\tau)}{d\tau}$ is unit-normalized,
$\dot{x}^\mu(\tau) \, \dot{x}_\mu(\tau) = -1$.  The quark's four-acceleration $\ddot{x}^\mu(\tau)$, being by definition normal to the 4-velocity, is then spacelike, so that 
\begin{equation}
a^2(\tau) \equiv \ddot{x}^\mu(\tau) \, \ddot{x}_\mu(\tau) \ge 0
\label{qaccel}
\end{equation}	
for any trajectory.

Given a quark following $x^\mu(\tau)$ with \mik\ boundary conditions, let us now specify its gravity dual.
Vacuum of SYM on Minkowski spacetime is described by
 (Poincare) AdS spacetime,
\begin{equation}
ds^2 = \frac{1}{u^2} \, \left( \eta_{\mu\nu} \, dx^{\mu}\, dx^{\nu} + du^2 \right)
\label{AdSPoincmet}
\end{equation}	
where we have WLOG set the AdS radius to unity, 
$\sqrt{\lambda}\, \alpha'  = 1$, so all distances are now measured in AdS units.  
The quark and its gluonic field is described by a fundamental string in \req{AdSPoincmet}
which ends on the boundary quark position $x^\mu(\tau)$ at $u=0$.
Mikhailov \cite{Mikhailov:2003er} solved the worldsheet equations of motion for the string with \mik\ boundary conditions, parameterized by the bulk radial coordinate $u$ and a time\footnote{
More specifically, $\tau$ is to the quark's proper time extended onto the worldsheet along ingoing null geodesics, thereby specifying the retarded time along the string worldsheet.
}
 coordinate $\tau$, and found a remarkably simple expression for the embedding coordinates $X^M(\tau,u) = (X^{\mu}(\tau,u) ,u)$:
\begin{equation}
X^{\mu}(\tau,u) = u \, {\dot x}^{\mu}(\tau) +  x^{\mu}(\tau) \ .
\label{WSsol}
\end{equation}	
The coordinate patch where this solution applies is specified by $u \in (0,\infty)$ and $\tau \in (-\infty, \infty)$; however as we shall see below, the induced metric on the worldsheet is geodesically incomplete.  In a snapshot at a constant $x^0$ slice of AdS, the string described by \req{WSsol} typically just ends in `mid-air'.

% - - - - - - - - - - - - - - - - -
\subsection{Worldsheet geometry}
\label{s:WS}
% - - - - - - - - - - - - - - - - -
To understand how much of the worldsheet is specified by \req{WSsol} let us  consider the two-dimensional worldsheet geometry.
The induced metric $\gamma_{ab}$ is easily obtained from \req{AdSPoincmet} and \req{WSsol},
\begin{equation}
\dsWS^2 = 
- \left( \frac{1}{u^2} - a^2(\tau) \right) \,  d\tau^2
- \frac{2}{u^2} \, d\tau \, du \ ,
\label{WSindmet}
\end{equation}	
where $a^2(\tau)$ is given in terms of the quark trajectory by \req{qaccel}. 
Lines of constant $\tau$ are null, while those of constant $u$ (given by orbits of ${\partial_\tau}$) are timelike near the boundary but can become spacelike deeper in the bulk, depending of $a$.
Note that distinct quark trajectories $x^\mu(\tau)$ can lead to the same induced metric on the worldsheet, as long as the contracted acceleration is the same; however, distinct functions $a^2(\tau)$ will lead to different worldsheets. 
In all cases, the string worldsheet 
is manifestly timelike, since its determinant is negative:
\begin{equation}
\det \gamma_{ab} = - \frac{1}{u^4} < 0 \ .
\label{WSmetdet}
\end{equation}	
Furthermore, for any regular quark trajectory $x^{\mu}(\tau)$, the worldsheet metric (\ref{WSindmet}) is manifestly regular for all $(\tau,u)$.
In fact, since $\gamma_{ab}$ has constant curvature $R^{(\gamma)}_{ab} = -\gamma_{ab} \ \ \Rightarrow \ \ R^{(\gamma)} = -2$,  it is a piece of AdS$_2$ embedded inside AdS$_5$.  This means that as such, the worldsheet geometry doesn't have any more refined structure.  However, when restricted to the Poincare patch (which itself is geodesically incomplete, being a proper subset of global AdS), the string worldsheet has several interesting coordinate singularities, namely at $u=0,\infty$ and $\tau =\pm \infty$.

\paragraph{Worldsheet coordinate singularities:}
There are four potential `edges' to the $(\tau, u)$ chart, we'll briefly indicate the geometrical meaning of each:
\begin{itemize}
\item $u=0$:  
In the full AdS$_{d+1}$ metric \req{AdSPoincmet}, $u=0$ is the boundary, i.e.\ any geodesic takes an infinite affine parameter to reach it.  Hence in the induced geometry  \req{WSindmet} on the worldsheet, $u=0$ likewise corresponds to the boundary (i.e.\ the set of infinitely distant endpoints for null and spacelike geodesics).
\item $u = \infty$:  
In the full geometry this corresponds to the Poincare horizon, so this is the locus of points (singularity from the 2-dimensional point of view) where the string exits the Poincare patch through the Poincare horizon.  
\item $\tau \to - \infty$:
This marks the future lightcone of the initial point along the quark's trajectory (when specified on Minkowski space, i.e.\ at $x^0 \to - \infty$).  Since by construction, no past-directed causal curve on the worldsheet can reach the boundary $u=0$ before that time, the $\tau = -\infty$ line represents the worldsheet past event horizon.  
In special cases, such as the trivial case of inertial quark with $a^2(\tau)=0$, this past event horizon coincides with the Poincare horizon; but in general it differs.  
\item $\tau \to \infty$:
This is the future lightcone of the future endpoint of the quark trajectory, however it is cutoff by $u=\infty$ and therefore does not enter in any interesting way.
\end{itemize}
We can now see that there are two distinct reasons the worldsheet can be geodesically incomplete: the Poincare horizon at $u=\infty$ and the past worldsheet horizon at $\tau = - \infty$.  The latter one is more interesting, as this set of points (for any positive finite $u$) is generically still within the Poincare patch.

To give a simple example, let us consider the case of constant acceleration, i.e.\ constant $a^2(\tau) = a^2$ (we will examine the uniform acceleration subclass in more detail in \sec{s:unifaccel}).
In this case the worldsheet metric \req{WSindmet} is static,
with $\partial/\partial \tau$ being the requisite Killing vector field, which is timelike at the boundary $u=0$, but becomes null at $u = u_0 \equiv 1/ a$.  In the globally static case of $a=0$, the worldsheet metric is precisely the Poincare patch of AdS$_2$, but in the more general case $ a^2>0$, the line element \req{WSindmet} can be thought of as a constant-angle slice of BTZ in ingoing Eddington coordinates, as can be verified by the transformation
$u = \frac{1}{w}$
combined with 
$\tau = T - \frac{1}{a} \, \tanh^{-1} \frac{w}{a}$
which gives
\begin{equation}
\dsWS^2 = - \left( w^2- a^2 \right) \,  dT^2
+ \frac{ dw^2}{w^2- a^2} \ .
\label{WSmetBTZ}
\end{equation}	
%

% Figure 
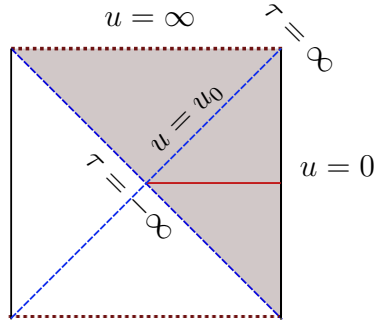
\begin{figure}
\begin{center}
\includegraphics[width=1.5in]{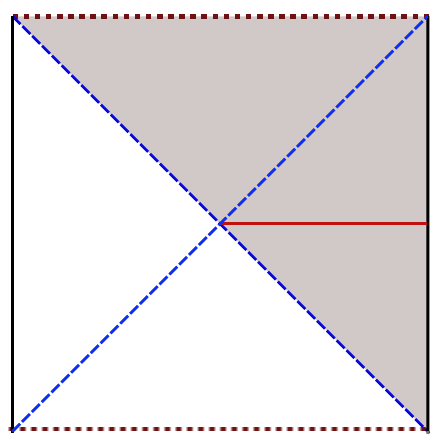}
\begin{picture}(0,0)
\setlength{\unitlength}{1cm}
\put(0,2){$u=0$}
\put(-2.6,4){$u=\infty$}
\put(-2,2.3){\rotatebox{45}{$u=u_0$}}
\put(-0.5,4.2){\rotatebox{-45}{$\tau =\infty$}}
\put(-2.9,2.2){\rotatebox{-45}{$\tau =-\infty$}}
\end{picture}
\caption{Penrose diagram for the worldsheet metric with constant ${\ddot x}^{\mu}(\tau) \,  {\ddot x}_{\mu}(\tau) = 1/u_0^2$.  The top and bottom horizontal dotted lines correspond to singularities at $u=\infty$ (i.e.\ Poincare horizon in the full AdS), the left and right vertical solid lines the $u=0$ AdS boundaries, and the dashed diagonal lines the worldsheet horizons.  The upper triangular shaded region corresponds to the full coordinate patch covered by the $(\tau, u)$ coordinates in the line element (\ref{WSindmet}).  The short red horizontal line is an example of a spacelike geodesic which terminates at finite affine parameter despite reaching $\tau = -\infty$.
}
\label{fig:WSPD}
\end{center}
\end{figure}

The latter form yields a more familiar context in which to ascertain which geodesics are incomplete.  The causal structure of the worldsheet metric  $\gamma_{ab}$ is indicated in the Penrose diagram of Fig.\ref{fig:WSPD}.  Since only the shaded part of the Penrose diagram is covered by the coordinates $(\tau,u)$, any geodesic which reaches the boundary of this region, while inside the full Penrose diagram, can be extended further. 
One example of such a geodesic
is the constant $T$ spacelike geodesic demarcated by the red segment in Fig.\ref{fig:WSPD} which corresponds to a snapshot of the string at constant  time $x^0=0$; another is the null horizon generator $\xi^a = \left( \frac{\partial}{\partial \tau} \right)^{\! a}$ (i.e.\ the upper portion of the $u = u_0$ curve in Fig.\ref{fig:WSPD}), since the affine parameter $\lambda$ along such null geodesic is related to $\tau$ by $\lambda = C \, e^{a \, \tau}$, which remains finite ($\lambda \to 0$) as $\tau \to - \infty$.

\paragraph{Worldsheet event horizon:}
Apart from the past horizon which marks the edge of the regime of validity of the worldsheet embedding \req{WSsol} and of the $(\tau,u)$ coordinate patch, there is another  special place on the string worldsheet, namely its (future) event horizon.  This is a globally-defined construct, signifying the boundary of the region on the worldsheet from which a signal can causally reach the boundary $u=0$ at some finite $\tau$.
For any smooth $a^2(\tau)\ge 0$ there is guaranteed to be a smooth worldsheet horizon, defined by the last null outgoing ray to reach infinity.  
 If $a(\tau)^2 =a^2$ is constant, then $u_0 = 1/a$ defines the worldsheet Killing horizon (where $\frac{\partial}{\partial \tau}$ becomes null), which for this static worldsheet coincides with the worldsheet event horizon.  However, for time-varying $a(\tau) \equiv \sqrt{a(\tau)^2}$ the event horizon is determined more globally, as we now explain.

Since the worldsheet metric is timelike,
 through any point there are two (i.e.\ left- and right-moving) uniquely-specified null curves (which must be null geodesics in the worldsheet metric, 
though not necessarily in the full spacetime\footnote{
When the string satisfies \mik\ boundary conditions, ingoing null geodesics on the worldsheet indeed do coincide with null geodesics of the ambient bulk geometry; however the outgoing ones typically do not.  For more general boundary conditions, generally neither set describes geodesics in the full spacetime, whereas on the two-dimensional worldsheet any null curve is of course automatically a null geodesic.}).
Let us for convenience use the coordinates  $\tau$  and $w$  defined above, in which the metric is simply
\begin{equation}
\dsWS^2 = - \left( w^2- a(\tau)^2 \right) \,  d\tau^2
+ 2\, d\tau \, dw \ .
\label{WSmetatau}
\end{equation}	
To find the `last null geodesic to reach the boundary' we need to solve
\begin{equation}
\frac{dw}{d \tau} = \frac{1}{2} \, \left( w(\tau)^2 - a(\tau)^2 \right)
\label{}
\end{equation}	
for the past-directed null geodesics emanating from large $w$ at late $\tau$.   

% Figure 
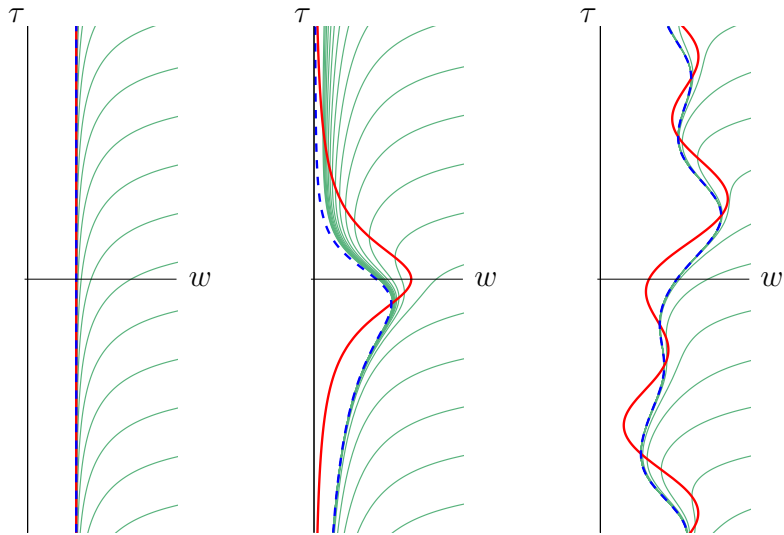
\begin{figure}
\begin{center}
\includegraphics[width=.8in]{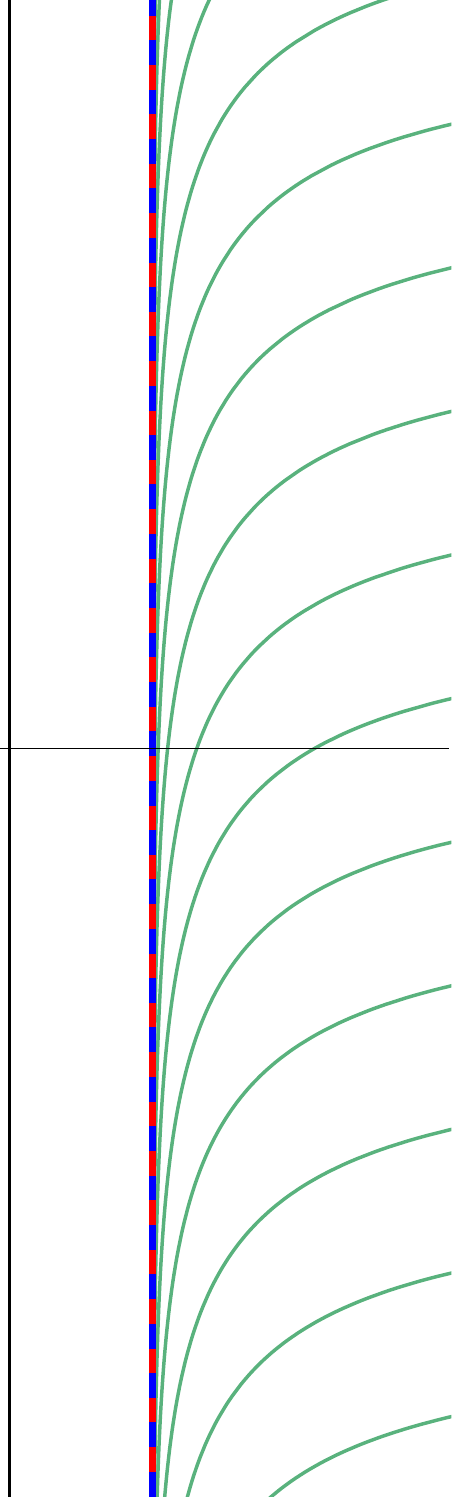}
\hspace{1.5cm}
\includegraphics[width=.8in]{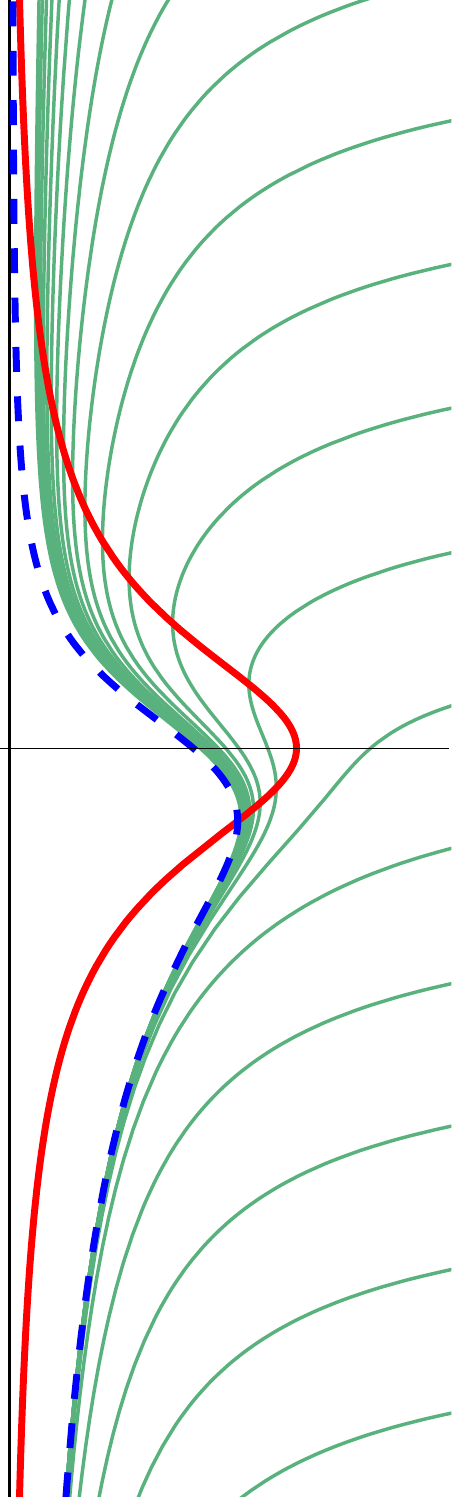}
\hspace{1.5cm}
\includegraphics[width=.8in]{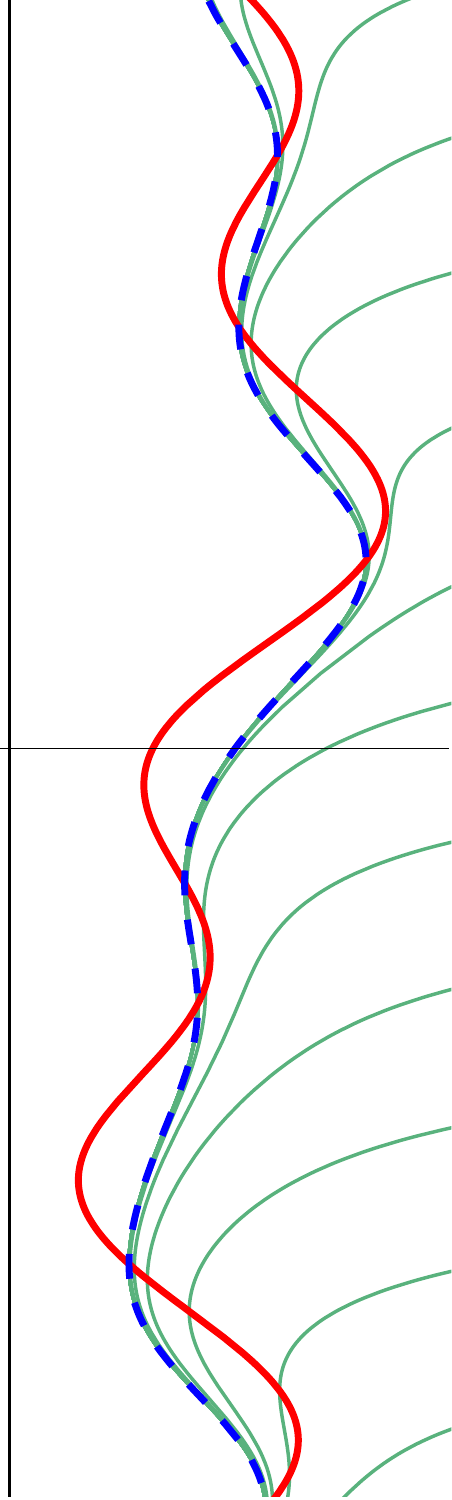}
\begin{picture}(0,0)
\setlength{\unitlength}{1cm}
\put(-0.,3.3){$w$}
\put(-3.8,3.3){$w$}
\put(-7.6,3.3){$w$}
\put(-10,6.8){$\tau$}
\put(-6.2,6.8){$\tau$}
\put(-2.4,6.8){$\tau$}
\end{picture}
\caption{Event horizons for spacetimes (\ref{WSmetatau}) for various accelerations $a(\tau)$ of the quark's trajectory.  The red curve indicates the acceleration $a(\tau)$, the thin green curves radial null geodesics, and the thick dashed blue curve the event horizon.
From left to right, we have chosen the acceleration function $a(\tau) = 1$, 
$a(\tau) = \frac{2}{1+ \tau^2}$, and $a(\tau) = 1+\sin^2 \tau +\sin \frac{\tau}{2} \, \cos \frac{\tau}{3}$.}
\label{fig:dynBHhor}
\end{center}
\end{figure}
In Fig.\ref{fig:dynBHhor} we plot the event  horizons for several 2-dimensional dynamical spacetimes (\ref{WSmetatau}) specified by a given $a(\tau)$.  
For orientation, we have plotted the acceleration curve $a(\tau)$ (red), several radial null geodesics $w(\tau)$, and the event horizon (dashed blue) which can be seen as the accumulation surface of the null geodesics.  We can see that even when the quark accelerates for a short time, the worldsheet has an event horizon.\footnote{
This result is not novel; for example, \cite{Chernicoff:2008sa}
pointed out that the process of energy loss due to quark's acceleration is accompanied by the formation of an event horizon on the corresponding worldsheet.
}  In general, the event horizon, being teleological, ``anticipates" the acceleration curve.
In other words, when $a(\tau)$ is some general function of $\tau$, we need to know it for arbitrarily late $\tau$ before we can solve for the event horizon -- that's what we mean by the worldsheet event horizon being globally defined.  

It is perhaps worthwhile  pointing out the following subtlety.
Although the teleological property of an event horizon being globally-defined holds for higher-dimensional black holes as well, there we can  at least recourse to quasi-local notions of horizon such as the apparent horizon which effectively captures the notion that locally gravity is so strong as to prevent even light from escaping.  In contrast,
 in two dimensions, there is {\it no} quasi-local indicator of the horizon whatsoever.
More precisely, there is no well-defined meaning of an apparent horizon (usually defined by marginally trapped surfaces, those with zero expansion along the outgoing null normal congruence), since there is no congruence -- the `surface' is a single point, and the `congruence' would be a single null curve from that point -- and correspondingly no notion of expansion. 
Looking at \req{WSindmet}, one might be tempted to think that the closest analog is the curve given by $u = 1/a(\tau)$ (given by the red curve in \fig{fig:dynBHhor}), which coincides with the worldsheet event horizon for constant acceleration, but this locus of points can be shifted by a mere coordinate transformation: there is nothing globally special about this curve.\footnote{
Note that in the $(\tau,u)$ coordinates it forms a 2-dimensional analog of an ergosurface (indeed, \cite{Chernicoff:2008sa} considered this `stationary limit curve' in the context of energy dissipation); however for generic $a(\tau)$ these coordinates have no geometrically-specified meaning.
}

Another way to see this is the following. Consider again the constant acceleration case.
As pointed out above, when the black hole is static, its event horizon coincides with the Killing horizon.  In higher dimensions, the bifurcation point is the fixed point of the Killing field, and corresponds to a co-dimension two extremal surface.  In two-dimensional static worldsheet, one could find a Killing field with fixed point anywhere we want; there is no transverse area which we could extremize.  To determine which Killing field is the relevant one, we need to use the global information of where does the string worldsheet encounter the Poincare horizon.  So a local observer sitting on the worldsheet would have no way of identifying where the horizon is.

% - - - - - - - - - - - - - - - - -
\subsection{String in global AdS}
\label{s:globalAdS}
% - - - - - - - - - - - - - - - - -
Having specified the geometry of a piece of the string worldsheet in the Poincare patch,
we are now ready to discuss the full completion of the string worldsheet in global AdS.  Of course, this task presumes that there is a natural way to globally complete the worldline of the quark itself.  We need to extend this worldline, originally specified on Minkowski spacetime by $x^\mu(\tau)$, beyond the Minkowski boundary.  How we do this influences how the rest of the worldsheet behaves, which will play an important role later, in \sec{s:resolcomp}.  Nevertheless, we'll see that for the worldlines of interest, such as that of quark undergoing inertial motion or uniform acceleration, there is a particularly `natural' way to smoothly extend the worldline to the full boundary ESU.
We start by indicating how to describe the known piece of string \req{WSsol} in global AdS, and then discuss its completion.

\paragraph{Coordinate transformation from Poincare to global AdS:}
The first task is quite straightforward and well-known.
To go between global and Poincare coordinates, we can use the embedding description of AdS$_5$ as a `hyperboloid' in $\RR^{4,2}$, and choose requisite embedding coordinates which manifest the different symmetries.
Global AdS$_{d+1}$ can be written as
\begin{equation}
ds^2 = - (r^2 + 1) \,  d\tg^2 + \frac{dr^2}{r^2+1} + r^2\, \left[ d\ph^2 + \sin^2 \ph \, d\Omega_{d-2}^2 \right]
\label{AdSglobal}
\end{equation}	
(the last term in square brackets merely expresses the $S^{d-1}$ metric $d\Omega_{d-1}^2$ in terms of the $S^{d-2}$ metric $d\Omega_{d-2}^2$ and the angle $\ph$ specifying the orientation of the Poincare patch).
The corresponding coordinate transformation between
the Poincare coordinates $\{x^\mu,u\}$ in which the metric is given by \req{AdSPoincmet}
and the global coordinates $\{\tg,r,\ph,\Omega^i\}$ in which the metric is given by \req{AdSglobal}
can then be written as 
\begin{equation}
u = \frac{1}{Q(t,r,\ph)} \ , \ \
x^0 = \frac{\sqrt{r^2 +1} \, \sin \tg }{Q(t,r,\ph)} \ , \ \
x^i = \frac{ r \, \sin \ph }{Q(t,r,\ph)} \, \Omega^i \ ,
\label{gtop}
\end{equation}	
where 
$\Omega^i$ denotes a unit vector on the $S^{d-2}$ and the denominator $Q$ is given by
\begin{equation}
Q(t,r,\ph) = \sqrt{r^2 +1} \, \cos \tg + r\, \cos \ph \ .
\label{gtopQ}
\end{equation}	
Note that the locus of points where $Q(t,r,\ph)=0$ marks the edge of the Poincare patch.
While the global coordinates cover the entire AdS spacetime (e.g.\ $r \to \infty$ only on the AdS boundary),
the Poincare patch of AdS specified by $u\in (0,\infty)$, and $x^\mu \in (-\infty,\infty)$ is geodesically incomplete since geodesics can reach the Poincare horizon $u=\infty$ in finite affine parameter.

\begin{figure}
\begin{center}
\includegraphics[width=6.25in]{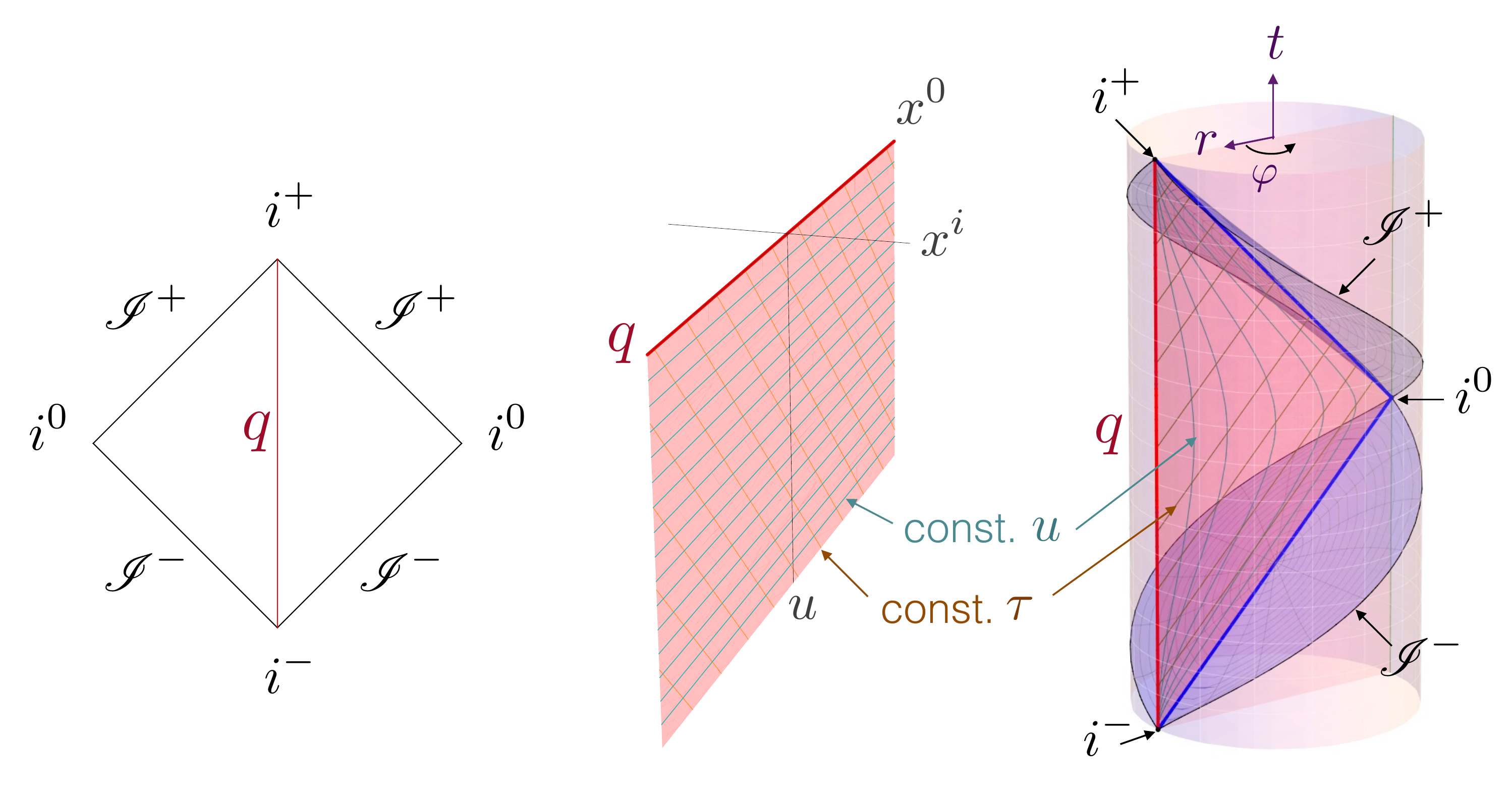}
\caption{
Static quark trajectory (red line, labeled by $q$) on boundary Minkowski Penrose diagram (left), with corresponding string worldsheet (red surface) in bulk on Poincare patch of AdS (middle) and global AdS (right);  in the latter the Poincare patch is indicated by the two null (purple) planes.  The string worldsheet coordinates $u$ and $\tau$ are shown by the faint (orange and cyan) lines as labeled, and the components of the (boundary) Minkowski spacetime boundary ($i^{\pm},i^0,\scri^\pm$ ) are as indicated.
}
\label{fig:staticq}
\end{center}
\end{figure}

\paragraph{Static quark example:}
To orient ourselves to the various sets of coordinates and notation, let us start with the simplest example, namely that of 
a static quark at the origin, $x^\mu(\tau) = (\tau,0,0,0)$. 
On the left panel of \fig{fig:staticq}, we plot this trajectory on the Penrose diagram\footnote{
As always, the Penrose diagram is obtained by a conformal rescaling of the spacetime which keeps radial null rays at 45 degrees, but contrary to convention we have unfolded the diagram (or equivalently indicated just a two-dimensional subspace) for ease of comparison with subsequent plots.
} of Minkowski spacetime.
We also label the various components of boundary of Minkowski spacetime, namely future and past timelike infinities $i^\pm$, future and past null infinities $\scri^\pm$, and spacelike infinity $i^0$, which comprise the endpoints of timelike, null, and spacelike geodesics, respectively.  Note that our static quark trajectory starts at $i^-$ and ends at $i^+$, which is generically the case unless the quark's acceleration lasts for infinite time (into the past or future).

% Figure 
The corresponding string worldsheet is given by
\begin{equation}
X^M(\tau,u) =  \left( u +\tau \ , 0 \ , \ 0   \ , \ 0 \ , \ u \right)   \ ,
\label{WSuv}
\end{equation}	
plotted in the middle and right panels of  \fig{fig:staticq} in Poincare and global AdS, respectively.  
Before discussing the worldsheet, let us briefly discuss the relation between these two representations.

As is conventional, we plot the radial $u$ coordinate of Poincare AdS as increasing vertically down, whereas the radial $r$ coordinate of global AdS increases radially outward (the AdS boundary indicated by the cylinder lies at $r=\infty$).  Note that the Minkowski Penrose diagram of the left panel, when bent into a cylinder,  matches correctly onto the AdS boundary in the global AdS representation of the right panel.  The past/future Poincare horizons (denoted by the lower/upper diagonal purple planes in global AdS) are then bounded by $(i^-,\scri^-,i^0)$ and $(i^0,\scri^+,i^+)$, respectively, and are generated by bulk null geodesics from $i^-$ and  $i^0$, which reconverge at $i^0$ and $i^+$, respectively.  In other words, the Poincare patch of AdS can be defined as the set of bulk points which are spacelike-separated from $i^0$.   The $u=\infty$ endpoint of the $u$ axis in the middle panel  corresponds to the boundary point $i^0$ in the right panel.  

For the static quark the string worldsheet in Poincare coordinates hangs `straight down', and in global coordinates stretches `straight across', but the entire infinite string worldsheet on the Poincare patch is captured by the (vertical triangle) piece in global AdS as indicated.  The rest of the global worldsheet is cutoff by the Poincare horizons, which the worldsheet intersects along $u=\infty$  (diagonal blue lines).
In this case 
 the string solution \req{WSsol} does not terminate abruptly inside the Poincare patch, because the past horizon $\tau = -\infty$ curve coincides with $u=\infty$.  However, in general (whenever the quark accelerates), this will no longer be the case; we will see an explicit example in \sec{s:unifaccel}. 

\paragraph{Global completion of string worldsheet:}

As we argued above, the worldsheet metric (\ref{WSindmet}) can be further extended.  
Correspondingly, a physical string does not just end in `mid-air' -- it continues beyond the coordinate patch.  
Since the string is a geometrical object (defined as the extremal surface with the requisite boundary conditions),  it doesn't care about which coordinates we describe it in, Poincare or global.  The \mik\ boundary conditions thus ensure that the worldsheet continues smoothly through the future Poincare horizon, because it is uniquely determined there by the quark trajectory.  For the case of static quark discussed above, this gives the top part of the vertical plane bisecting AdS (indicated by a fainter shading in \fig{fig:staticq}; the faint green line through $i^0$ then corresponds to the trajectory of the induced \aq).  

However, this only gives twice the extent of the original worldsheet \req{WSsol}, not the infinitely larger extent present in global AdS.  In particular, the \mik\ boundary conditions applied to quark trajectory specified by $x^\mu(\tau)$ do not determine the worldsheet in the past of the $\tau = -\infty$ endpoint ($i^-$ in this case) or the future of the  $\tau = +\infty$ endpoint ($i^+$).  To go beyond these, we need to supply further information about the quark trajectory on ESU.

In the static quark case viewed on global AdS (cf.\ right panel of  \fig{fig:staticq}), there is a very natural way to do this: since the quark trajectory is given by a vertical line at $r=\infty$, $\ph=0$, and $\tg \in (-\pi,\pi)$,  we simply need to extend it into the straight line ($r=\infty$, $\ph=0$) for all $t$.  In the bulk this then naturally extends the string worldsheet into the vertical strip bisecting the global AdS cylinder.  Its intersection with the boundary consists of two trajectories on ESU, one corresponding to the quark and the other to the \aq\ (in \fig{fig:staticq} indicated by the red and faint green vertical lines, respectively). 

While the static quark case is particularly simple, similar extension will work more generally; in fact the case of uniform acceleration considered in \sec{s:unifaccel} is virtually identical.
For a generic quark trajectory described by real analytic functions $x^\mu(\tg)$, there will be a unique analytic continuation extending these to the entire range $t \in (-\infty,\infty)$.\footnote{
Of course, if we allow non-analytic trajectories for the quark (for example if we abruptly force the quark in a new direction) then there can be infinitely many extensions.
}

% - - - - - - - - - - - - - - - - -
\subsection{Criterion for induced anti-quark}
\label{s:aqcriterion}
% - - - - - - - - - - - - - - - - -

Now that we have established a convenient way for examining the string worldsheet corresponding to a quark on arbitrary trajectory in flat space (which as we discussed can be extended into a trajectory on the ESU boundary of global AdS), we are in a position to revisit our main puzzle:  Under which circumstances does the string return to the boundary so as to induce an \aq\ within a single Poincare patch?  The  answer to this question 
 rests on what knowledge the quark trajectory gives us about the other end of the corresponding string worldsheet.  As discussed above, this is determined by the \mik\ boundary conditions, so we first turn to these.

Since viewing the worldsheet on global AdS has the virtue of allowing us to easily keep track of both endpoints of the string, let us  consider the implication of the \mik\ boundary conditions in this language.  As pointed out already in \cite{Mikhailov:2003er}, the effect of the quark's behavior at a given $\tau$ influences the field theory configuration only along the future lightcone.  Null geodesics on the ESU reconverge on the antipodal point of the $S^3$, at $\Delta \tg = \pi$ (in AdS units) time later, so they imply the presence of the \aq\ at that point.  The bulk analog of this statement is that all bulk null geodesics (one of which will coincide with the worldsheet ingoing null geodesic) from a given point likewise reconverge at the antipodal point $\Delta \tg = \pi$ time later.  For example, in the static quark case considered above, the fact that the quark appeared at $i^-$  forces the \aq\ to pass through $i^0$.  

This simple observation is actually the crux in formulating the criterion for the presence of the induced \aq.  Whenever the quark on Minkowski spacetime starts out at $i^-$, the \aq\ is forced to pass through $i^0$, on a timelike trajectory (since the quark is timelike).  Causality then implies that it cannot pass through the remainder of the Poincare patch, the entirety of which is spacelike-separated from $i^0$.   In other words, a quark trajectory of this type cannot induce an \aq. Under what conditions would  the quark trajectory  {\it not} start at $i^-$?  This can happen only when the quark is past-eternally accelerating, so that its `initial' velocity approaches the speed of light.  In such a case its trajectory in Minkowski spacetime will start out at a starting point of a light ray, i.e.\ somewhere along $\scri^-$ on the corresponding Penrose diagram.  A well-known example of such a case is the uniform acceleration one, examined in more detail in \sec{s:unifaccel}.  It is easy to see that in this case, the antipodal null-related point to the quark's starting point must lie along $\scri^+$ rather than $i^0$.  But if the \aq\ passes through $\scri^+$, then (again by causality) it must have passed though its past, i.e.\ through the {\it interior} of the Minkowski spacetime forming the boundary of the same Poincare patch.  

We have argued that past-eternal acceleration is a necessary criterion for the trajectory to start at $\scri^-$, but is it a sufficient one?  Clearly, a circular trajectory for all time such as considered in \cite{Athanasiou:2010pv}\footnote{
See also e.g.\ \cite{Chernicoff:2010yv,Hubeny:2010bq,Chernicoff:2011vn} for more detailed discussions of the string worldsheet.}
 would {\it not} start from $\scri^-$ since it does not approach a light ray.
 Indeed, any spatially bounded trajectory, no matter what its acceleration $a(\tau)$, necessarily starts at $i^-$ rather than at $\scri^-$.
The circular case corresponds to a constant-magnitude (and hence `eternal' in the sense of not falling off at early/late times) acceleration, but not constant direction.  This example makes it clear that we need to strengthen our criterion to rule out such cases.  Any uni-directional acceleration with magnitude which is bounded from below would suffice, but it is not necessary:  The acceleration could easily change directions with the quark's trajectory nevertheless remaining spatially unbounded and beginning somewhere on $\scri^-$ (which is tantamount approaching a light ray trajectory in the remote past).\footnote{
Hence our criterion is actually more general than the criterion specified in e.g.\ \cite{Garcia:2012gw} of the trajectory approaching {\it uniform} acceleration in the  past.
}  
For lack of more compact convenient phrasing, we will then state our criterion simply as follows.

\paragraph{Induced \aq\ criterion:} 
{\it A quark following a given trajectory on flat space with \mik\ boundary conditions will induce an \aq\ iff its trajectory approaches that of a light ray in the past.}  This means that the quark has a past Rindler horizon, which simultaneously forms the future Rindler horizon for the \aq\ from the other side.

%____________________________________________
\section{Uniformly accelerating quark-\aq\ pair}
\label{s:unifaccel}
%____________________________________________

Having formulated a criterion for when the presence of a quark in Minkowski spacetime with \mik\ boundary conditions naturally induces a corresponding \aq, let us now turn to the simplest and best-known example of this curious phenomenon, namely that of the quark undergoing uniform acceleration.
Physically, such a scenario may be achieved by the presence of a constant electric field; we discuss the SYM construction of this setup and its regime of validity in more detail in 
 \cite{HSpap1}; 
here we simply take the trajectory as our starting point and focus on the behavior of the corresponding classical string worldsheet. 

For concreteness, consider a quark moving in the $x^1$ direction with constant unit\footnote{
It is easy to describe any other constant acceleration $a$ as well, which for completeness we do in \App{s:UAaMink}, and can in fact be related to any other $a$ by a symmetry of AdS, but  $a = 1$ is particularly convenient given our choice of coordinates.
} acceleration, such that at early/late times it asymptotes to an ingoing/outgoing light ray through the origin.
We can describe the quark's worldline on Minkowski spacetime by 
\begin{equation}
x^{\mu}(\tau) =  \left( \sinh \tau \ , \ 
\cosh \tau \ , \ 0 \ , \ 0 \right) 
\label{qrkWLuA}
\end{equation}	
which is simply the right branch of the hyperbola 
$-(x^0)^2 + (x^1)^2 = 1$.
Equivalently, one can express the trajectory more straightforwardly in light cone coordinates
(defined by $x^\pm \equiv x^1 \pm x^0$) as 
$x^\pm(\tau) = e^{\pm \tau}$.
% Figure 
\begin{figure}
\begin{center}
\includegraphics[width=2.6in]{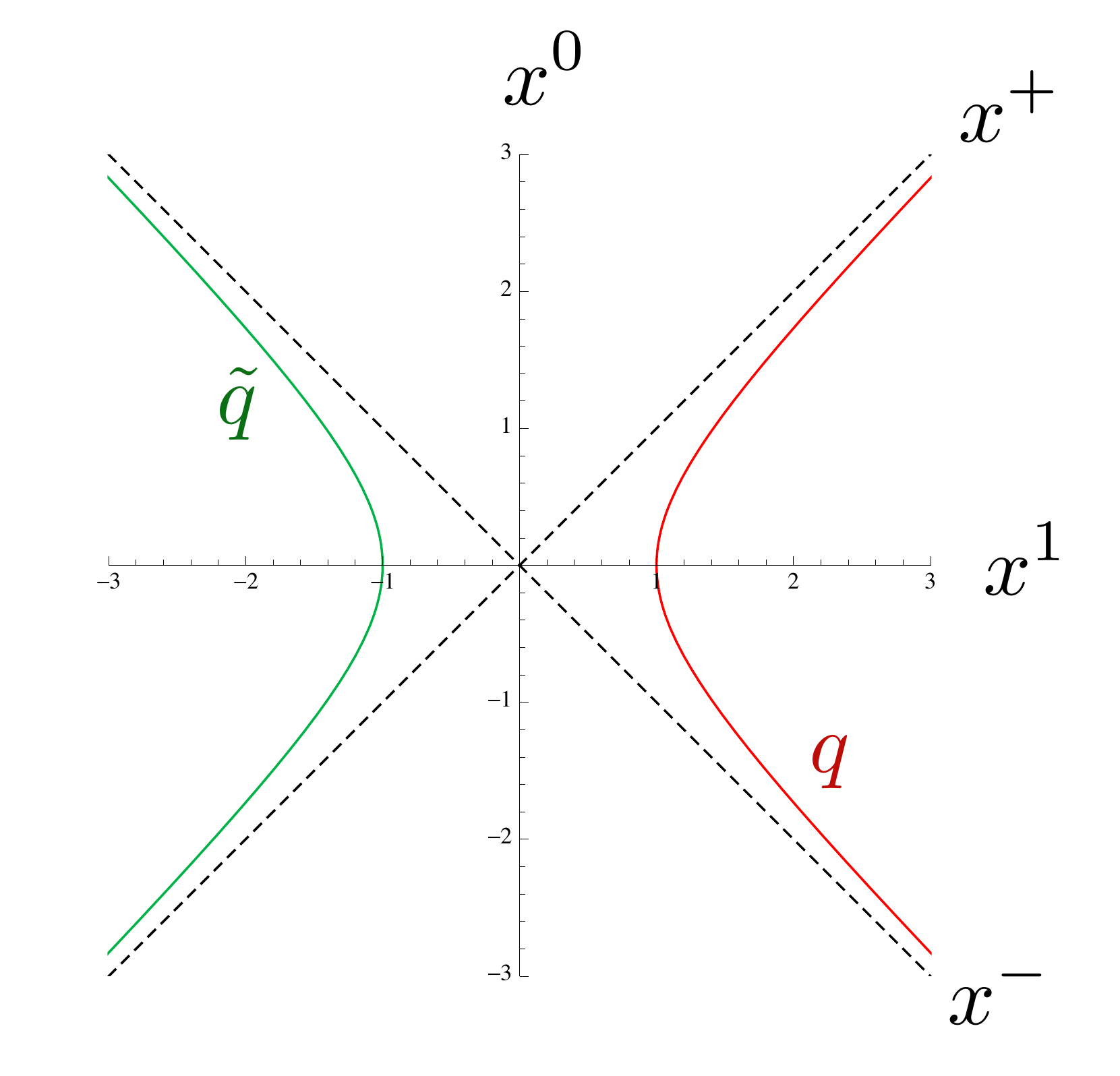}
\hspace{1cm}
\includegraphics[width=2.6in]{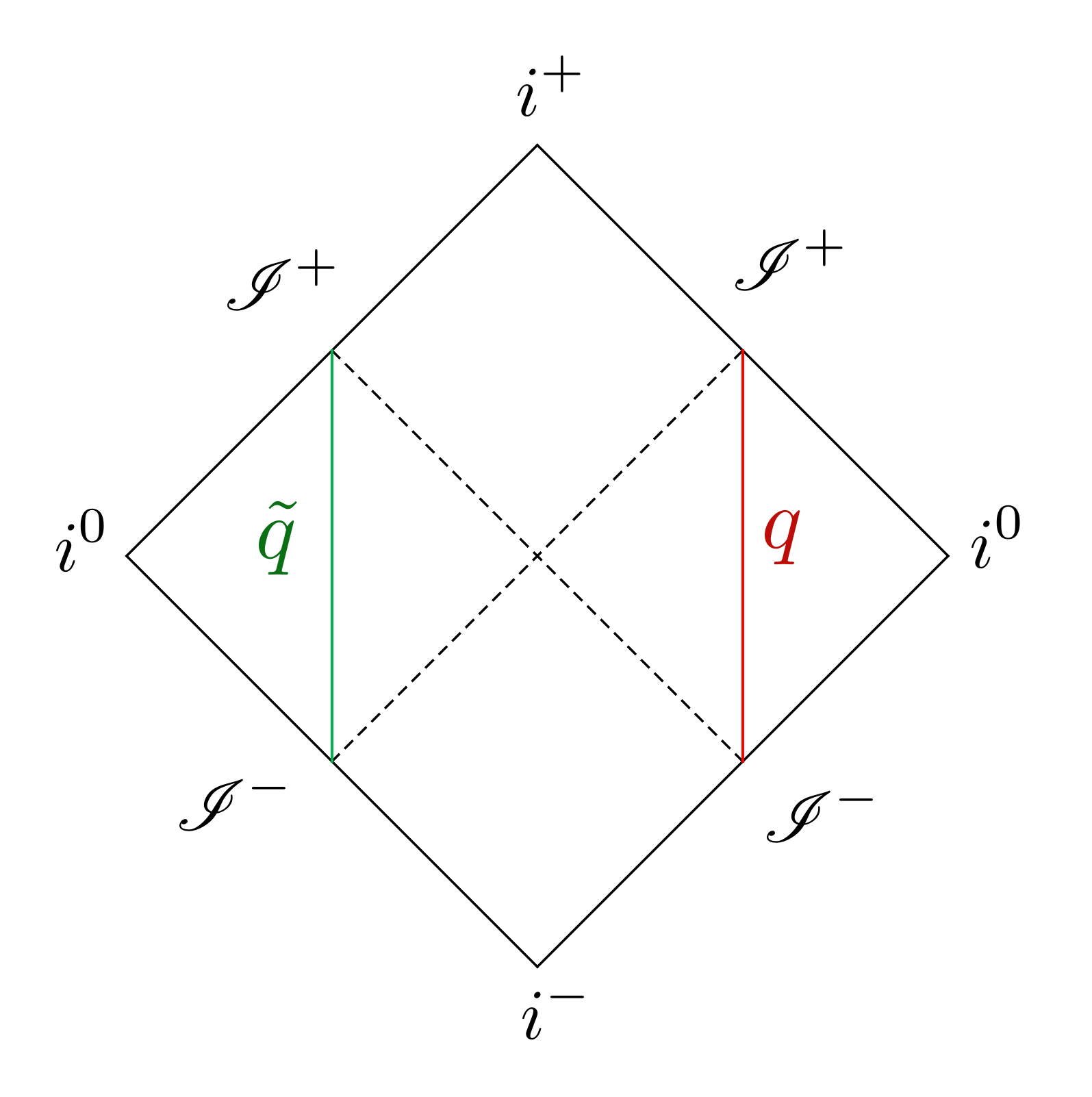}
\caption{
Quark ($q$, red) on uniformly-accelerating trajectory (\ref{qrkWLuA}), and its induced \aq\ (${\tilde q}$, green), on Minkowski spacetime, plotted in Cartesian coordinates (left) and on Penrose diagram (right).  The lightcone coordinate axes $x^\pm$ are denoted by the diagonal dashed lines, which form the Rindler horizons for the quark and anti-quark.  Note that the (anti-)quark trajectory can end on $\scri^\pm$ because it is eternally accelerating and therefore not following a geodesic. 
}
\label{fig:qaqtraj}
\end{center}
\end{figure}
In \fig{fig:qaqtraj} we indicate the quark trajectory by the red curve (labeled $q$) in standard Cartesian coordinates $x^\mu$ (left) and on the Penrose diagram (right).
Note that in contrast to the static case, now the quark starts form $\scri^-$ and ends on $\scri^+$ of Minkowski spacetime, and therefore it has Rindler horizons (dashed diagonal lines \fig{fig:qaqtraj}), which are simply the lightcone axes $x^\pm$.  In other words, there is a region of Minkowski spacetime $\{x^\pm<0 \}$ which remains causally disconnected from the quark trajectory $\{x^\pm>0 \}$; it can neither influence nor be influenced by the quark.  As we will explain below based on the string worldsheet, this is precisely the region traversed by the induced \aq\ (green curves labeled ${\tilde q}$ in \fig{fig:qaqtraj}).

\begin{figure}
\begin{center}
\includegraphics[width=6.in]{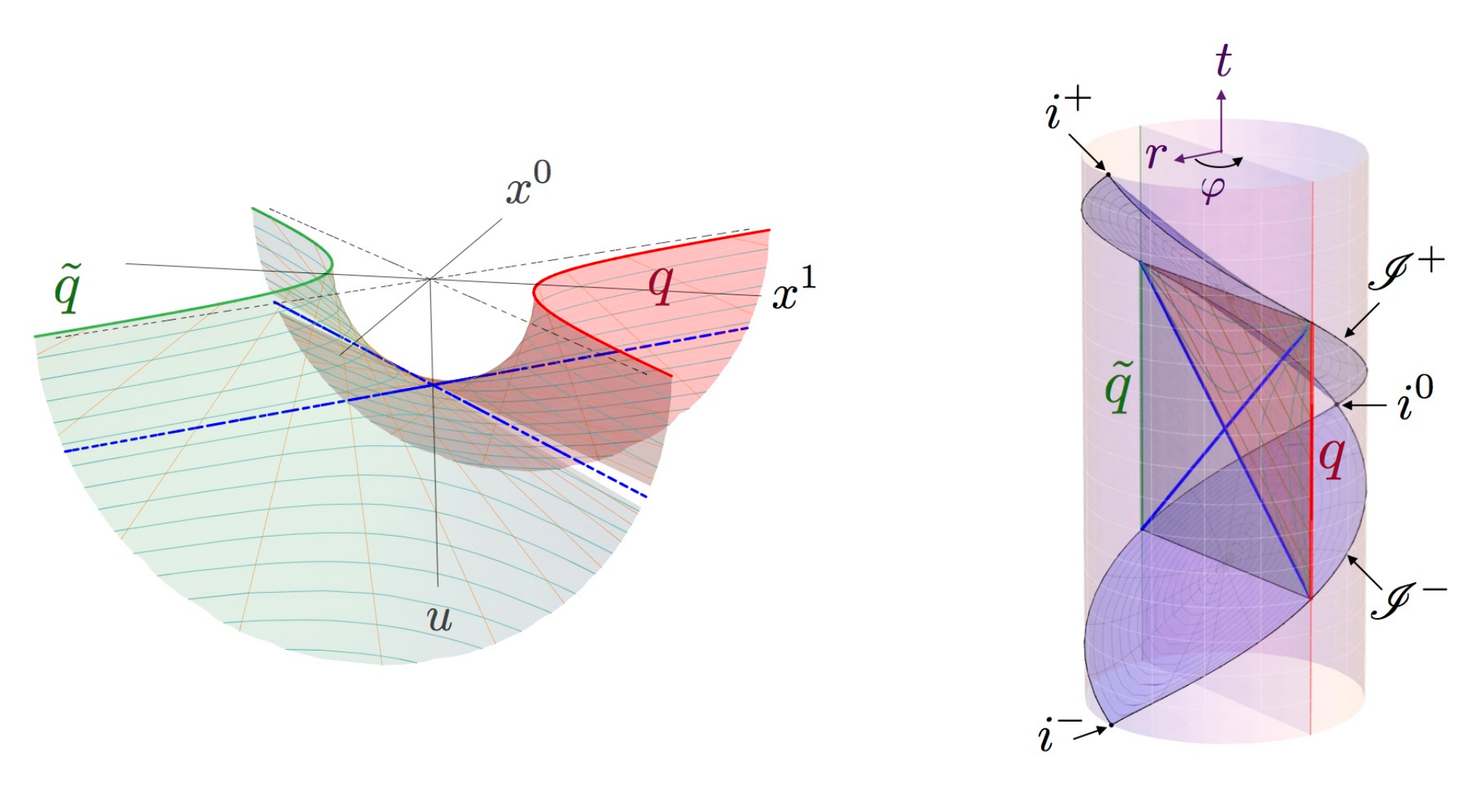}
\caption{
String worldsheet (red surface) for the uniformly-accelerating quark (red curve) and its global extension (green surface) which induces a `mirror' \aq\ (green curve), plotted both in Poincare-AdS (left) and in global AdS (right).  The quark and \aq\ trajectories are as depicted in Fig.\ref{fig:qaqtraj}; 
 The green surface is the completion of the worldsheet, which can be equivalently obtained by time-reverse of \req{WSsol} applied to the anti-quark.  The surface mesh depicts lines of constant $\tau$ (orange) and $u$ (cyan) as in \fig{fig:staticq}.  The past and future worldsheet event horizons (blue dashed lines) both lie at $u=1$.}
\label{fig:qaqWS}
\end{center}
\end{figure}

The string worldsheet corresponding to the quark trajectory \req{qrkWLuA}, embedded into Poincare AdS$_5$ as determined by  \req{WSsol}, is given by
\begin{equation}
X^M(\tau,u) = \left( u \, \cosh \tau +\sinh \tau \ , \ 
u \, \sinh \tau + \cosh \tau  \ , \ 0   \ , \ 0 \ , \ u \right)   \ .
\label{WSuA}
\end{equation}	
This is indicated by the red surface in \fig{fig:qaqWS}, which shows the string worldsheet on Poincare AdS (left) and global AdS (right), analogously to the middle and right panels of \fig{fig:staticq}.
% Figure 
We see that in Poincare AdS, near the boundary the worldsheet `hangs down' from the quark, but because of the acceleration, it also trails behind.  In fact, a snapshot of the string at constant Poincare time $x^0$ shows that for $x^0>0$, the string actually bends back out towards the boundary.  Though it remains at $u=1$ when $x^+=0$, relative to the overall depth to which the string penetrates, this bending becomes increasingly more pronounced at late times.
But the crucial point is that the worldsheet \req{WSuA} ends abruptly in the bulk.  This occurs as $\tau \to - \infty$, at the past worldsheet horizon (indicated by the right-slanting dashed blue line in \fig{fig:qaqWS}).

Note that the worldsheet embedding satisfies the relation
\begin{equation}
-(x^0)^2 + (x^1)^2 = x^+ \, x^- =  1- u^2 \ ,
\label{WSsurfuA}
\end{equation}	
though \req{WSsol} is valid only for $x^+ > 0$ (corresponding to $\tau \in (-\infty, \infty)$). 
Nevertheless, the surface \req{WSsurfuA} is well defined for all $x^+$.  This gives a unique analytic extension of the string worldsheet to negative $x^+$, indicated by the green surface in \fig{fig:qaqWS}.  This surface reaches back out to the boundary at the other branch of the hyperbola, which induces the `mirror' \aq, undergoing a unformly accelerated trajectory, but in the opposite direction, as shown.

Indeed, the extension is even more manifest in global AdS, where the string worldsheet is easy to guess.  Since the quark trajectory on the Penrose diagram (cf.\ \fig{fig:qaqtraj}) is a straight vertical line, the corresponding string worldsheet in global AdS must be a flat plane (i.e.\ drawn as a vertical strip), just as it was for the static quark of \fig{fig:staticq}.  (Indeed, the two cases are related by a global rotation $\ph \to \ph + \pi/2$.)
The $\tau \to - \infty$ curve slices this flat plane diagonally as indicated in right panel of  \fig{fig:qaqWS}, so the obvious extension of the worldsheet is the continuation of this plane to the other side of AdS.

Despite the global rotational symmetry which relates the worldsheet for uniformly accelerating quark-\aq\ pair to that of a single static quark, the causal structure of the worldsheet metric is very different from the static case, since the relative positioning of the worldsheet with respect to the Poincare patch differs.  In particular, the worldsheet now intersects the Poincare horizons along constant $\tg$ (given by the thin black horizontal lines in right panel of \fig{fig:qaqWS}), rather than along a null curve.  Correspondingly, the worldsheet has a nontrivial causal structure, with event horizon (indicated by the diagonal blue lines); cf.\ \fig{fig:WSPD}.
The worldsheet completion within the full Poincare patch then has another asymptotic region (whose boundary describes the other endpoint of the string), as well as a past singularity (along the past Poincare horizon).

To summarize, this simple example nicely illustrates the criterion formulated in \sec{s:aqcriterion}:  because the quark undergoes eternal acceleration, the \aq\ induced by the other endpoint of the corresponding string worldsheet is necessarily present on the Minkowski spacetime.  The worldsheet causal structure then describes an eternal black hole with two asymptotic boundaries.\footnote{
This structure was recently noted \cite{Jensen:2013ora} 
 to support the `ER=EPR' conjecture of \cite{Maldacena:2013xja}, since the quark-\aq\ singlet is entangled (cf.\ also \cite{Sonner:2013mba}, generalized more substantially in \cite{Chernicoff:2013iga}).  However, as pointed out above, the Einstein-Rosen `bridge' in this case is present solely due to the global structure of the worldsheet -- there is no local `wormhole neck' present.} 

%____________________________________________
\section{Avoidance of induced anti-quark}
\label{s:resolution}
%____________________________________________

We have seen that under innocuously natural conditions on the quark trajectory in Minkowski spacetime (namely whenever its acceleration does not fall off fast enough at early times or average to zero) and boundary conditions (namely when only outgoing radiation, sourced by the quark's acceleration, is present), the corresponding bulk dual predicts a curious effect:  the quark's trajectory {\it induces} an accompanying \aq!  This is true even in the absence of Gauss law constraint on Minkowski background, where the SYM a-priori allows color charged states.
One may then wonder whether the string theory poses a nontrivial constraint on the allowed processes in the strongly coupled SYM that have not yet been understood from the field theory side directly, or whether one can circumvent the above conclusion by some other means.  

As previewed in the Introduction, we propose the latter resolution.  In fact, there are multiple ways of avoiding the presence of the \aq, which we revisit in \sec{s:resolcomp}; from the global standpoint the most natural one is to modify the boundary conditions, the subject of \sec{s:bentWS}.

% - - - - - - - - - - - - - - - - -
\subsection{Globally static smooth worldsheet}
\label{s:bentWS}
% - - - - - - - - - - - - - - - - -

In this section we present the bulk dual to a single eternally accelerating quark on Minkowski spacetime with no accompanying \aq. 
As argued in \sec{s:aqcriterion}, this is not possible if the dual string worldsheet responds only to the quark `source' via the \mik\ boundary conditions.  The obvious solution then is to modify our boundary conditions.  From the global standpoint, where nothing particularly singles out one endpoint of the string from the other, it seems most natural to impose Dirichlet boundary conditions at both endpoints.  The worldsheet can then straddle global AdS in a far vaster variety of ways, specified by {\it two} arbitrary timelike boundary trajectories rather than just one.   

If we restrict attention to a single Poincare patch of global AdS so as to describe the boundary configuration on Minkowski background rather than on the full ESU, we can arrange for the absence of the \aq\ simply by forcing its trajectory on ESU to pass through the $i^0$ of the Poincare patch.  No matter what the rest of the trajectory looks like, this guarantees that the \aq\ will not appear anywhere within the Minkowski background corresponding to the boundary of the given Poincare patch of AdS.  
The disadvantage of this construction is that unlike the \mik\ boundary condition case reviewed in \sec{s:setup}, generally there is no closed-form expression for the worldsheet embedding in AdS in terms of the two endpoint trajectories.  In particular,  the string worldsheet is no longer given by the  simple expression \req{WSsol}. Instead one has to solve the boundary-value problem on a case-by-case basis, and explicit solutions can be found only for configurations with sufficient symmetry.  

However, for the case of greatest interest, namely that of the uniformly accelerating quark considered in \sec{s:unifaccel}, there does exist a closed-form solution for the worldsheet if we take the \aq\ endpoint to follow a globally static trajectory.\footnote{
In fact, since the string worldsheet is globally static and flip-symmetric for a family of uniformly-accelerating trajectories with unspecified  acceleration, we can easily find the solution for this more general case.  For completeness, we present this class of solutions in \App{s:UAaDir}.
}
We will now examine this solution in some detail,  as it illustrates the main physical points.
Let us consider the quark trajectory \req{qrkWLuA}, smoothly extended to the full ESU.  In global AdS coordinates the quark is static: it sits at constant $\ph = \phq \equiv \pi/2$ for all $\tg$. We can then take the other endpoint of the string worldsheet to likewise remain static in global coordinates, i.e.\ to sit at constant $\ph=\phaq =\pi$ so that it passes through $i^0$.  With these boundary conditions, the entire string worldsheet is  static, which renders the solution tractable.

Since the worldsheet profile in global AdS coordinates is independent of $\tg$, we can characterize its embedding by a single function $\ph(r)$ (reflected smoothly and symmetrically around its deepest point $\rmin$ such that the endpoints at $r\to \infty$ reach $\phq$ and $\phaq$).  
This solution is already known, but for completeness, in \App{s:UAaDir} we construct it explicitly for arbitrary $\phq$ by solving the equations of motion obtained from the Nambu-Goto action, i.e.\ by extremizing the full classical string worldsheet area.
Although the actual expression for $\ph(r)$, given in \req{phiofrL}, \req{phiofrR}, is unilluminating, its derivative is quite simple:
\begin{equation}
\ph'(r)^2 = \frac{\rmin^2 \, (\rmin^2 +1)}{r^2 \, (r^2 +1)\, \left[ 
r^2 \, (r^2 +1) - \rmin^2 \, (\rmin^2 +1)\right]} \ ,
\label{phiintgrnd}
\end{equation}	
where the deepest reach $\rmin$ is fixed in terms of $\phq$; for the case of unit acceleration $\phq \equiv \pi/2$, we have $\rmin \approx 0.62$.

% Figure 
\begin{figure}
\begin{center}
\includegraphics[width=2.5in]{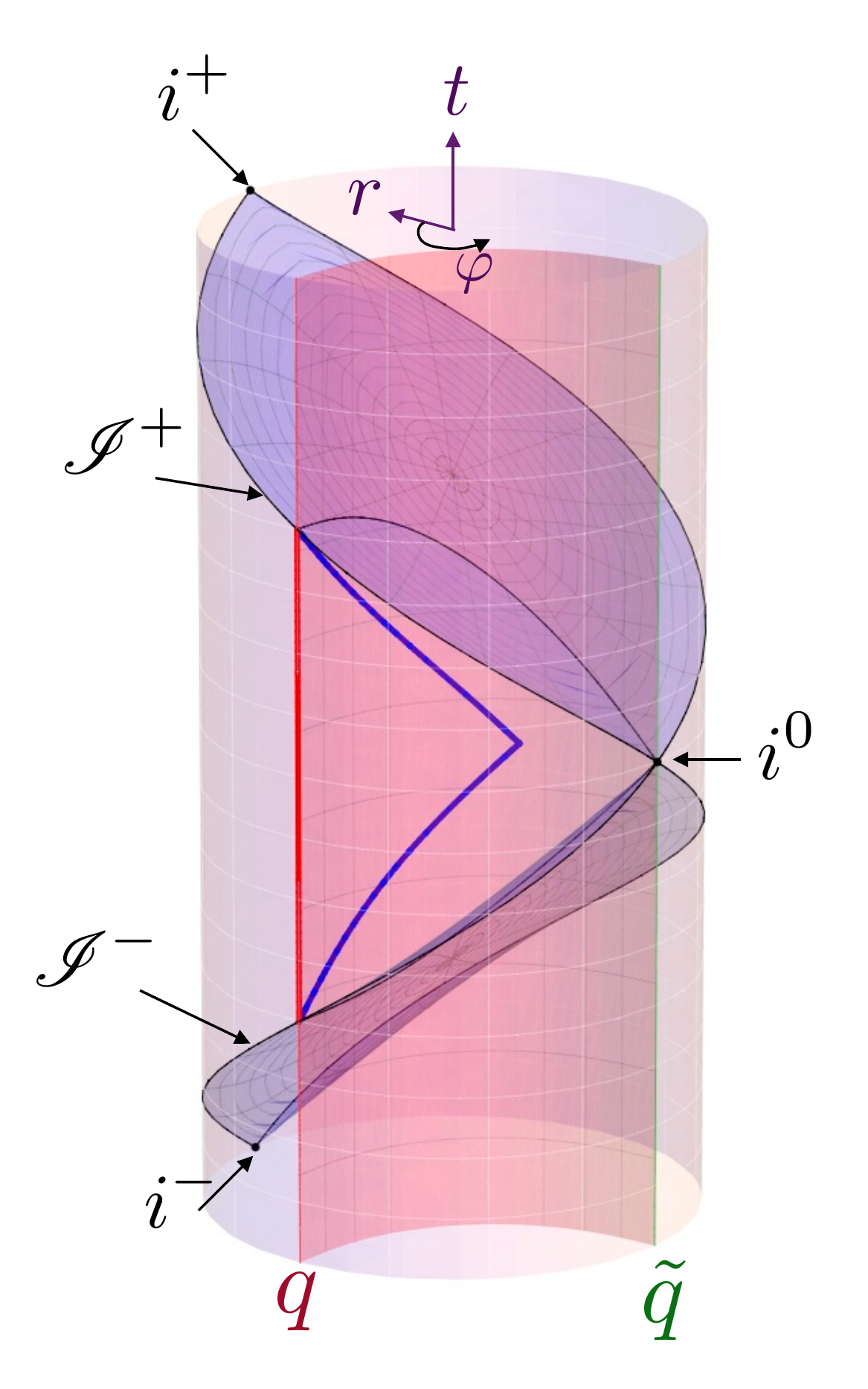}
\caption{
String worldsheet, plotted in global AdS, for a single uniformly accelerating quark (red line labeled $q$).  The anti-quark (faint green line labeled ${\tilde q}$) is forced to pass through $i^0$ and therefore does not appear in the original Minkowski space.  
Poincare wedge, its intersection with the worldsheet, and the worldsheet horizons are shown with the same conventions as in the right panel of \fig{fig:staticq} and \fig{fig:qaqWS}.
}
\label{fig:WSglobalForced}
\end{center}
\end{figure}
In \fig{fig:WSglobalForced} we plot the string worldsheet in global coordinates (red surface).  This is directly comparable with the right panels of \fig{fig:staticq} and \fig{fig:qaqWS}, except that we have rotated the plot so as to display the worldsheet better.
Instead of bisecting AdS straight through, the worldsheet is now bent so as to pass through $i^0$.  
The worldsheet metric $\gamma_{ab}$ is of course static, 
\begin{equation}
\dsWS^2 = 
- (r^2 + 1) \, d\tg^2
+ \frac{r^2}{r^2 \, (r^2 +1) - \rmin^2 \, (\rmin^2 +1)} \, dr^2
\label{}
\end{equation}	
but it is no longer just a piece of AdS$_2$.  In particular, although we of course still have $R^{(\gamma)}_{ab} = -\gamma_{ab}$ as before, the Ricci scalar (explicitly computed in \req{bentWSRicci}) now has a mild dependence on $r$, being most curved around its deepest reach and asymptoting to the requisite AdS value at the boundary.
It is also evident that the worldsheet is no longer ruled by null geodesics (since all null geodesics in AdS necessarily reconverge at the antipodal point of the boundary sphere).

Let us now focus on just the Poincare patch of AdS (bounded by the purple null planes from $i^0$ in \fig{fig:WSglobalForced}).  Its boundary then corresponds to Minkowski spacetime, where by construction, only the quark end of the string is visible.  
Although the worldsheet is globally static, this is not manifest in the Poincare coordinates  \req{AdSPoincmet}.  Indeed, the Poincare horizon gives rise to two distinguished pairs of curves on the worldsheet: its intersection with the Poincare horizon (indicated by the black curves in \fig{fig:WSglobalForced}), and its event horizon (indicated by the blue curves as before).  The former are two spacelike curves which meet at $i^0$ and correspond to coordinate singularities ($u=\infty$).
Although there is only one timelike asymptotic boundary present on the Penrose diagram of this part of the string worldsheet (unlike the double-sided case in \sec{s:unifaccel}), the fact that these curves are spacelike automatically implies that the worldsheet (restricted to the Poincare patch) has nontrivial causal structure: it has a black hole.

The worldsheet event horizon is again generated by null curves on the worldsheet which end at the quark trajectory where it intersects the Poincare horizon.
The actual expression for null geodesics on this background can be written in terms of Elliptic functions, and is relegated to the Appendix, cf.\ \req{bentWShor}.
It is easy to show that in general the location of the `bifurcation' point of this event horizon lies beyond the point of flip symmetry at $\rmin$; indeed in the case of vanishingly small acceleration $\phq \to 0$, the bifurcation point recedes to $i^0$, thus recovering the causally trivial Poincare patch of AdS$_2$.

To summarize, we have seen that it is easy to engineer a bulk configuration whose dual, when viewed on the boundary of a single Poincare patch, corresponds to a single uniformly accelerating quark on Minkowski background with no accompanying \aq.  In fact, there are infinitely many such ways -- all that is required is for the \aq\ trajectory to pass through $i^0$ -- but here we have chosen a particularly simple one which renders the solution explicitly tractable.  Nevertheless, the resulting configuration is not as `nice' as the quark-\aq\ pair considered in \sec{s:unifaccel} (as manifested, for example, by the worldsheet not being ruled by null geodesics), which is directly related to the fact that the system no longer satisfies the \mik\ boundary conditions.  
In fact, the SYM side reflects this as well, as seen e.g.\ in \cite{HSpap1}: the delicate cancellations in the perturbative calculations found for the quark-\aq\ pair (which allowed us to obtain an exact result which matches a version of the disc amplitude in the string theory dual), does not occur in this case, rendering analogous computation intractable.

While modifying the boundary conditions seems to us the most natural means to achieve the desired end (namely a single quark on any chosen trajectory), in \sec{s:resolcomp} we contrast this with other possibilities that have previously been proposed in the literature.

% - - - - - - - - - - - - - - - - -
\subsection{Comparison with previous resolutions}
\label{s:resolcomp}
% - - - - - - - - - - - - - - - - -

In \sec{s:unifaccel} and \sec{s:bentWS} we have presented two physically distinct configurations involving a uniformly accelerating quark on Minkowski spacetime, with correspondingly distinct bulk duals.  Let us now take stock of the larger set of possibilities, given the quark trajectory \req{qrkWLuA}.
From the bulk perspective, at best only part of the string worldsheet is determined by $x^{\mu}(\tau)$; there are several natural ways to complete the worldsheet, each having some advantages along with some drawbacks.\footnote{
We thank Alberto Guijosa for useful discussions of  these issues.}

To fix terminology, let us consider the following 4 setups, in historical order:
\begin{enumerate}
\item[a)]  Mikhailov \cite{Mikhailov:2003er}: Single uniformly accelerating quark, with incomplete string worldsheet \req{WSsol}; given by the red part of the worldsheet depicted in \fig{fig:qaqWS}.
\item[b)] Xiao \cite{Xiao:2008nr}:  Uniformly-accelerating quark-anti-quark pair given by the natural smooth completion of \req{WSsol} beyond Mikhailov's $(\tau,u)$ coordinate patch; depicted by the red plus green parts of the worldsheet in  \fig{fig:qaqWS}.
\item[c)] Garcia, Guijosa, Pulido \cite{Garcia:2012gw}:  Worldsheet satisfying Mikhailov's boundary conditions but nevertheless allowing only a single quark on the Minkowski boundary, by modifying quark's initial conditions; as discussed below, the corresponding string worldsheet has null parts and is not smooth.
\item[d)] Present paper:  Smooth string worldsheet which does not satisfy Mikhailov's boundary conditions as discussed in \sec{s:bentWS}; depicted in \fig{fig:WSglobalForced}.
\end{enumerate}

Setup a), described by the solution \req{WSsol} found by  Mikhailov \cite{Mikhailov:2003er} using \mik\ boundary conditions, is all that can be determined within the given Poincare patch of AdS.  
Nevertheless, it is clear that this can't represent the complete bulk dual, since the string  would just end abruptly at a finite distance.  
In contrast, Xiao's \cite{Xiao:2008nr} smooth extention of the worldsheet everywhere within the Poincare patch (setup b) happilly avoids this unphysical situation.  However, since the worldsheet bends so much that it attains the boundary at the other end as well, this implies the presence of a mirror \aq.  That in turn is surprising and potentially undesirable from the field theory perspective, where we expect to be able to consider a single accelerating quark in isolation.

It is important to note that the smooth extension used by \cite{Xiao:2008nr} implicitly assumed the behaviour of the quark before the beginning  of (Poincare $x^0$) time, as well as \mik\ boundary conditions.
To avoid the \aq's presence, then, we need to relax one of these assumptions: either consider different boundary conditions, or use different extension of the string worldsheet (corresponding to modifying the initial conditions).  We have chosen the former approach (setup d), as described in \sec{s:bentWS}, sacrificing the \mik\ boundary condition.

The authors of \cite{Garcia:2012gw} (setup c), on the other hand, proposed the latter resolution, further discussed in \cite{Chernicoff:2013iga}, tantamount to modifying the initial configuration of the gluonic field.  Instead of modifying the boundary conditions on the string worldsheet away from the \mik\ one, they effectively modify the fact of eternal acceleration.  One can consider a limiting case of trajectories $x^{\mu}(\tau)$ wherein the quark was moving at a constant velocity $v$ close to the speed of light initially, and then started uniformly accelerating.  In our set-up, this would require the electric field to be switched on at some finite time approaching $\tau \to -\infty$.  On the Minkowski spacetime Penrose diagram, such a trajectory would start from $i^-$ and therefore the antipodal trajectory must indeed pass through $i^0$ -- so there is no mirror \aq, essentially by construction.  The scenario of \cite{Garcia:2012gw} has this property.

The price that one pays for this is the strangeness of the string worldsheet in the $v \to c$ limit:  it is in effect composed of two pieces, one null and the other timelike, joined at a kink, where the string has a discontinuous normal.\footnote{ 
On the other hand, this resolution ameliorates one of puzzles mentioned in \cite{HSpap1} (cf.\ also \cite{Garcia:2012gw}) related to what part of the string worldsheet contributes to the quark's energy:  just the part above the event horizon or everywhere.  While the former is well-motivated for situations with well-defined Euclidean continuation (since the Euclidean `black hole'  caps off smoothly at the horizon), in general it seems rather too global a construct to encode as (temporally) local quantity as the energy.  If a part of the worldsheet is null, it doesn't contribute to the area, so we can include it or not as convenient.  Unfortunately, however, the worldsheet of \cite{Garcia:2012gw} remains timelike past the its {\it future} event horizon, where this `resolution' does not apply.}
 On the boundary, the limiting null quark trajectory would require either infinite energy source or a massless quark initially, moving at the speed of light.  Hence while seemingly more innocuous and physically motivated, we find this `resolution' stranger than the one we advocate in \sec{s:bentWS}.  In our case, the string is smooth everywhere and well-behaved from the global viewpoint, but the price that {\it we} pay is the modification of the (apparently natural) boundary conditions.
  
To summarize, each of the setups a)-d) mentioned above, involving a quark undergoing eternal acceleration, have some bizarre feature.  In case a), the string worldsheet is incomplete and just ends abruptly.  In case b), we're forced to consider a mirror anti-quark even in absence of Gauss law constraint.  In case c), we have to contend with kinky string worldsheet corresponding to part of the quark trajectory being genuinely lightlike.  Finally, in our case, d), while avoiding all these drawbacks, we sacrifice the naturalness of the boundary conditions.

%____________________________________________
\section{Discussion}
\label{s:discussion}
%____________________________________________

Innocent as the setup may seem, the story of eternally accelerating quark is a rich and puzzling one.  Specification of the quark trajectory does not fully determine the complete system:  one must supplement the worldline $x^{\mu}(\tau)$ by boundary conditions on the gluonic fields as well.  Said differently, the entanglement structure of the system plays a crucial role.  From the bulk perspective, only part of the string worldsheet is determined by $x^{\mu}(\tau)$; there are several natural ways to complete the worldsheet, as outlined in \sec{s:resolcomp}, each having some advantages along with some drawbacks.  In this context, the AdS/CFT correspondence has a rather remarkable implication:  the seemingly innocuous field theory scenario corresponding to a single eternally accelerating quark on Minkowski background with purely outgoing gluonic field profile an no shock wave is {\it inconsistent}.

We have seen that, for visualization purposes, it is extremely useful to consider the location of the string in global coordinates, as this demystifies many of the naively bizarre aspects.  
For example, the incompleteness of the string worldsheet given by  \req{WSsol} mentioned above is more manifest: in Poincare AdS we only see the incompleteness at the past horizon ($\tau = -\infty$), whereas in global coordinates we also see the incompleteness at the Poincare horizons ($u = \infty$).  
Moreover, assuming the worldsheet is smooth, it is a straightforward exercise to write the full geodesically complete extension of the string worldsheet in global AdS.  
This has suggested a natural generalization of the worldsheet boundary conditions to describe a single uniformly accelerating quark with no accompanying \aq, as discussed in \sec{s:bentWS}.

Although we have focused on computationally tractable examples with large amount of symmetry in order to illustrate the main physical points explicitly, the lessons which can be gleaned are quite general.  Assuming the \mik\ boundary conditions for the bulk string, it is easy to show that for {\it any} quark trajectory which is {\it eternally} accelerating in the past so as to approach a light ray, there will be an accompanying anti-quark induced on the $\RR^{3,1}$ boundary spacetime.  While we have exemplified this effect in \sec{s:unifaccel} using the unique smooth extension of the string worldsheet \req{WSsol}, {\it any} timelike extension of the quark trajectory $x^\mu(\tau)$ beyond $\tau= -\infty$ would lead to the same conclusion.  This is because the \mik\ boundary conditions guarantee that the  presence of a quark at $\scri^-$ induces an \aq\ at the antipodal point on $\scri^+$, and therefore in the past of $\scri^+$ -- i.e.\ the interior of the Minkowski spacetime.

If we wish to avoid the presence of the induced \aq, we need to either modify the boundary conditions for the string (i.e.\ relinquish the \mik\  boundary conditions), or the initial conditions for the quark (i.e.\ relinquish the quark actually entering the spacetime through $\scri^-$, on what in global AdS boundary would be a timelike trajectory).  In our discussion in \sec{s:bentWS}, we chose the former:  we specified the trajectories for {\it both} string endpoints, forcing the other one to pass through $i^0$ of Poincare patch, and thereby precluding its appearance in the Minkowski boundary spacetime. 
For the case of uniform acceleration, the entire string worldsheet can in fact be made globally static, but for any quark trajectory whatsoever, there are infinitely many ways (trajectories for \aq\ on ESU) in which to achieve the absence of \aq\ in flat space.

It is a valid question to what extent modifying the boundary conditions was a physically natural resolution.  Indeed, Mikhailov's \mik\ boundary conditions would seem more physically relevant, since signals only propagate `down' the string, i.e.\ away from the endpoint, which in the boundary corresponds to radiation emanating from the accelerating quark and none coming in from infinity.  
However, since these seemingly natural boundary conditions are at odds with the naive expectation that we should be able to consider a single quark in isolation on any timelike trajectory, we have to enlarge our considerations.   The proposed Dirichlet boundary conditions pose a well-defined and physically sensible system, and therefore are in a sense equally valid.

\paragraph{Note added:}  As this paper was nearing completion, we learned of \cite{Alberto} which has some overlap with the discussion in \sec{s:resolcomp}.

%%%%%%%%%%%%%%%%%%%%%%%%%%%%%%%%%%%____________________________________________
\subsection*{Acknowledgements}
\label{acks}
%____________________________________________
It is a pleasure to thank Rajesh Gopakumar, Alberto Guijosa, and Mukund Rangamani for useful discussions. 
G.W.S thanks NSERC of Canada for support.
VH was supported in part by the Ambrose Monell foundation, and by the STFC Consolidated Grant ST/J000426/1.

\appendix
%____________________________________________
\section{String worldsheet for uniformly accelerating quark}
\label{s:arbacc}
%____________________________________________

In the main text we have considered uniform {\it unit} ($a(\tau)=1$) acceleration for simplicity in order to illustrate the basic point.  However, since it is easy to generalize to unspecified uniform acceleration $a$, we present the corresponding string solutions here for completeness.

% - - - - - - - - - - - - - - - - -
\subsection{Solution for \mik\ boundary conditions}
\label{s:UAaMink}
% - - - - - - - - - - - - - - - - -

Let us first generalize the setup of \sec{s:unifaccel} to arbitrary uniform acceleration.
This is the easy case of \mik\ boundary conditions, where the full worldsheet in Poincare AdS is given by \req{WSsol}.
If the quark is moving in the $x^1$ direction with constant acceleration $a$,
its worldline on Minkowski spacetime is given by 
\begin{equation}
x^{\mu}(\tau) = \frac{1}{a} \, \left( \sinh(a \, \tau) \ , \ 
\cosh(a \, \tau)  \ , \ 0 \ , \ 0 \right) 
\label{qrkWLuAa}
\end{equation}	
(or $x^\pm(\tau) = \frac{1}{a} \, e^{\pm a \, \tau}$ in light cone coordinates),
which generates the right branch of the hyperbola $-(x^0)^2 + (x^1)^2 = \frac{1}{a^2}$.

The string worldsheet corresponding to the quark trajectory \req{qrkWLuA}, embedded into Poincare AdS$_5$ is then given by
\begin{equation}
X^M(\tau,u) = \left( u \, \cosh(a \, \tau) +\frac{1}{a} \, \sinh(a \, \tau)  \ , \ 
u \, \sinh(a \, \tau) +\frac{1}{a} \, \cosh(a \, \tau)  \ , \ 0   \ , \ 0 \ , \ u \right)   \ .
\label{WSuAa}
\end{equation}	
Plotted in global AdS, only the $a=1$ case looks like a flat plane bisecting global AdS; in all other cases the worldsheet looks bent.  Nevertheless, the intersection with the Poincare horizon is fixed for all cases, and the worldsheet Penrose diagrams looks just like for the $a=1$ case.

% - - - - - - - - - - - - - - - - -
\subsection{Solution for Dirichlet boundary conditions}
\label{s:UAaDir}
% - - - - - - - - - - - - - - - - -

Now let us turn attention to the generalization of the more interesting solution presented in \sec{s:bentWS}, corresponding to a single uniformly accelerating quark (with no accompanying \aq) with unspecified acceleration $a$.
Although the string worldsheet is not given by the simple expression  \req{WSsol}, we can nevertheless find an explicit solution for the string profile thanks to the symmetry  of the problem.  In particular the worldsheet is specified by a single function $\ph(r)$ (reflected smoothly and symmetrically around its deepest point $\rmin$ such that the endpoints at $r\to \infty$ reach $\phi_q$ and $\pi$.  

%-----------------------
\paragraph{Worldsheet profile:}

Consider global AdS$_{d+1}$, with one angular direction $\ph$ singled out, as given by \req{AdSglobal}.  
To find the solution for the string worldsheet in this spacetime, we extremize its area.
Let us make the gauge choice for the worldsheet coordinates to be  $\sigma^a=(\tg,r)$.  Denoting $' \equiv \frac{\partial}{\partial r}$, we calculate the induced metric $\gamma_{ab}$ on the worldsheet in terms of the spacetime metric $g_{\mu\nu}$ and coordinates $X^{\mu}$ in the usual way,
$\gamma_{ab} = g_{\mu\nu} \, \partial_a X^{\mu} \, \partial_b X^{\nu}$,
finding the determinant 
$-\gamma \equiv -\det \gamma_{ab} = 
1+r^2 \, (r^2+1) \, \ph'(r)^2$. 
One can extremize the area of this surface,
$A = \int \sqrt{-\gamma} \, d^{2}\sigma $
by solving the corresponding Euler-Lagrange equations
with 
\begin{equation}
{\cal L}(\ph(r),\ph'(r);r)=\sqrt{1+r^2 \, (r^2+1) \, \ph'(r)^2}
\label{wslag}
\end{equation}	
whose solution can be obtained by integration:
\begin{equation}
\frac{\partial}{\partial r} \, \frac{\partial {\cal L}}{\partial \ph'} = \frac{\partial {\cal L}}{\partial \ph} =0
\qquad \Rightarrow \qquad
\frac{\partial {\cal L}}{\partial \ph'} 
= \frac{r^2 \, (r^2+1) \, \ph'(r)}{\sqrt{1+r^2 \, (r^2+1) \, \ph'(r)^2}}
= C
\label{}
\end{equation}	
where $C$ is a constant determined by the boundary conditions:  Denoting the minimal radius reached by $\rmin$, we note that at the turning point $\ph'(\rmin) = \infty$, which sets
\begin{equation}
C=\rmin \, \sqrt{\rmin^2 +1} \ .
\label{}
\end{equation}	
This allows us to solve for $\ph'(r)$,
\begin{equation}
\ph'(r)^2 = \frac{\rmin^2 \, (\rmin^2 +1)}{r^2 \, (r^2 +1)\, \left[ 
r^2 \, (r^2 +1) - \rmin^2 \, (\rmin^2 +1)\right]}
\label{phiintgrnda}
\end{equation}	
and integrate to obtain $\ph(r)$ as an Elliptic Integral of the Third Kind.
Since the worldsheet is double-valued in $r$, we get two branches 
\begin{equation}
\ph_L(r)
= \pi + \Re\left[ \frac{1}{\sqrt{\rmin^2+1}} \, 
\Pi \left( \frac{2 \, \rmin^2+1}{\rmin^2+1} \ , \ 
\sin^{-1} \sqrt{\frac{r^2+ \rmin^2+1}{2 \, \rmin^2+1}}
 \ , \  2+ \frac{1}{\rmin^2} \right) 
 \right]
\label{phiofrL}
\end{equation}	
and 
\begin{equation}
\ph_R(r) = 2 \pi+2 \, \ph_m- \ph_L(r) \ ,
\label{phiofrR}
\end{equation}	
which corresponds to the reflection around the angle of symmetry,
\begin{equation}
\ph_m 
 = \frac{\phq+\pi}{2}
\equiv \ph_L(\rmin)
= \pi +  \Re\left[  \frac{1}{\sqrt{\rmin^2+1}} 
 \Pi \left( \frac{2 \, \rmin^2+1}{\rmin^2+1} \ , \ 
 2+ \frac{1}{\rmin^2} \right) \right] \ .
\label{}
\end{equation}	
Note that the total separation between the endpoints is given by 
$\Delta \ph 
= 2 \, (\pi - \ph_m) = \pi - \phq$.
In order to have $\Delta \ph = \pi$ (so that the string stretches straight across), $\rmin=0$ as expected, whereas in order to have $\Delta \ph = \pi/2$ (corresponding to the case of unit-uniformly accelerating quark with no antiquark in Minkowski boundary spacetime), $\rmin \approx 0.62451$.
Curiously, to within a few percent (and better at large and small $\rmin$), $\phq(\rmin)$ is very well approximated by
\begin{equation}
\phq(\rmin) \approx 2 \, \tan^{-1} \frac{\pi \, \rmin}{2}
\qquad \Leftrightarrow \qquad 
\rmin(\phq) \approx  \frac{2}{\pi} \, \tan \frac{\phq}{2} \ .
\label{}
\end{equation}	
%

% Figure 
\begin{figure}
\begin{center}
\includegraphics[width=6.4in]{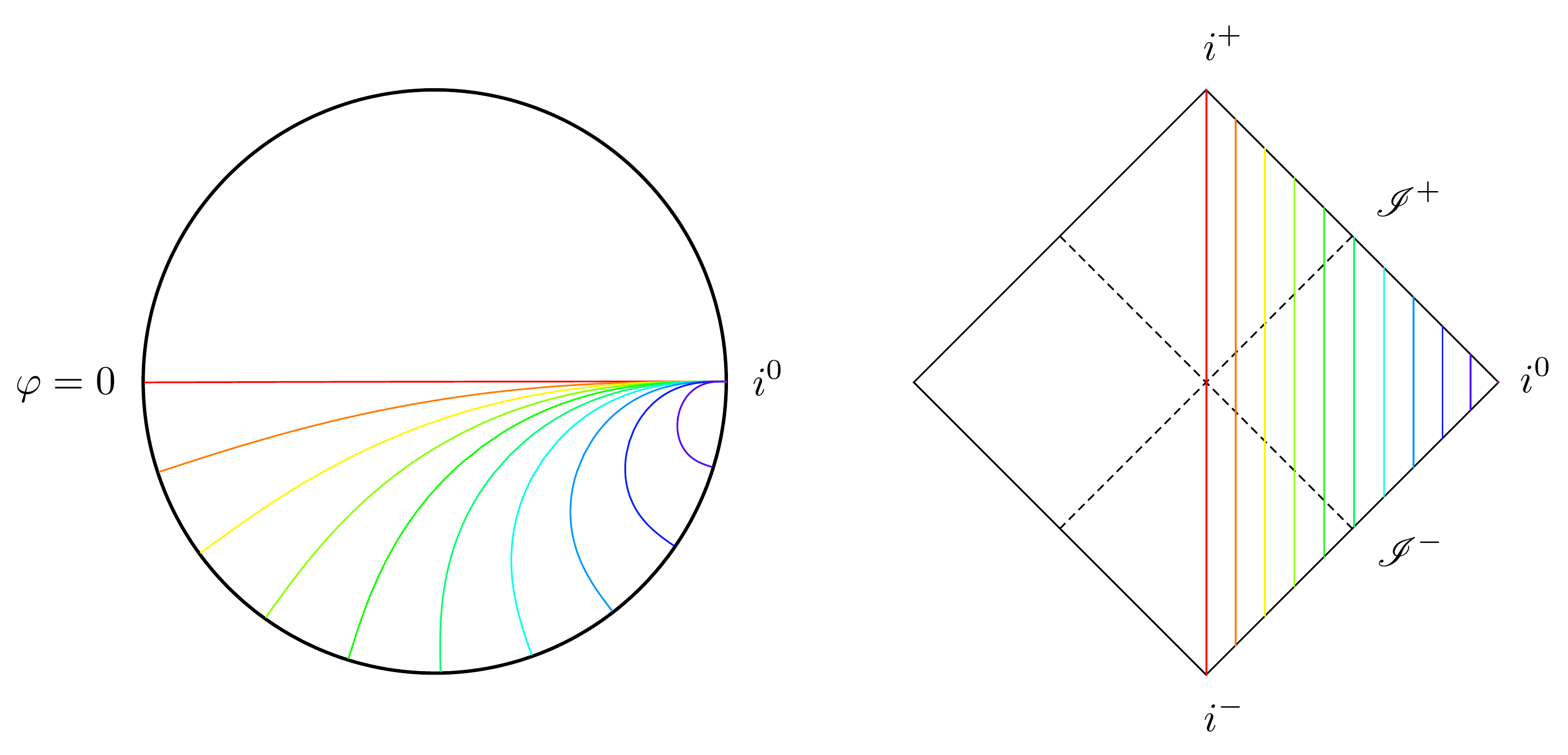}
\caption{
(Left): Profile of the string worldsheet for various endpoints, plotted on Poincare disk (constant $\tg$ slice of global AdS; cf.\ Fig.\ref{fig:WSglobalForced} looked at from below), for the single uniformly accelerating quark, with the anti-quark held at $i^0$.  AdS boundary is represented by the thick black circle and the various worldsheet profiles (color-coded by the proximity of string endpoints, or relatedly quark acceleration) are given by the thin curves.
(Right):  The corresponding quark trajectories on Minkowski spacetime Penrose diagram (cf.\ right panel of Fig.\ref{fig:qaqtraj}).  
The static case and unit acceleration case discussed above are both special cases in this family, depicted by the upper/left (red) lines and the middle (green) curves, respectively.
}
\label{fig:WSPoincDiskForced}
\end{center}
\end{figure}

\begin{figure}
\begin{center}
\includegraphics[height=3in]{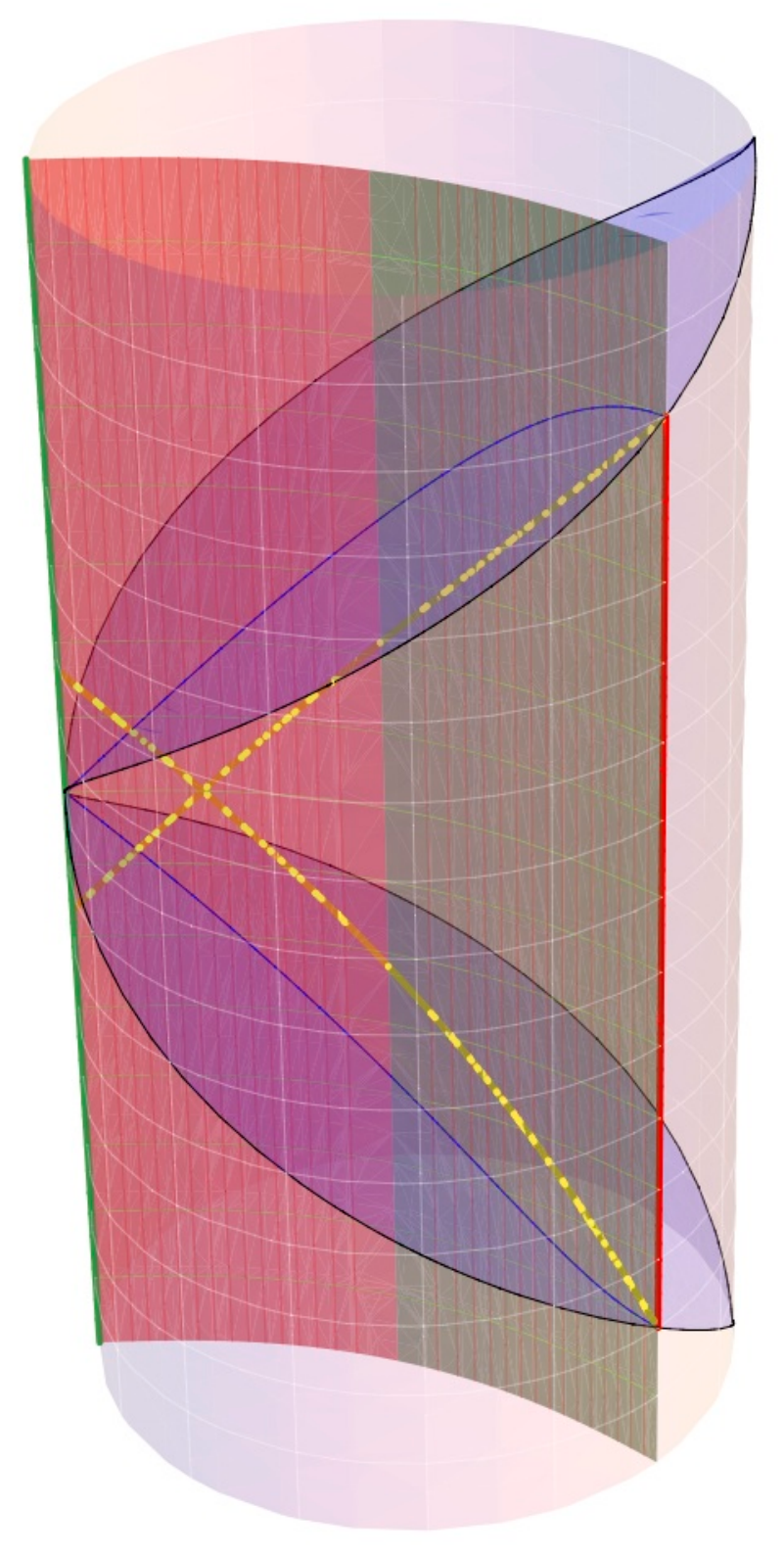}
\hspace{1cm}
\includegraphics[height=3in]{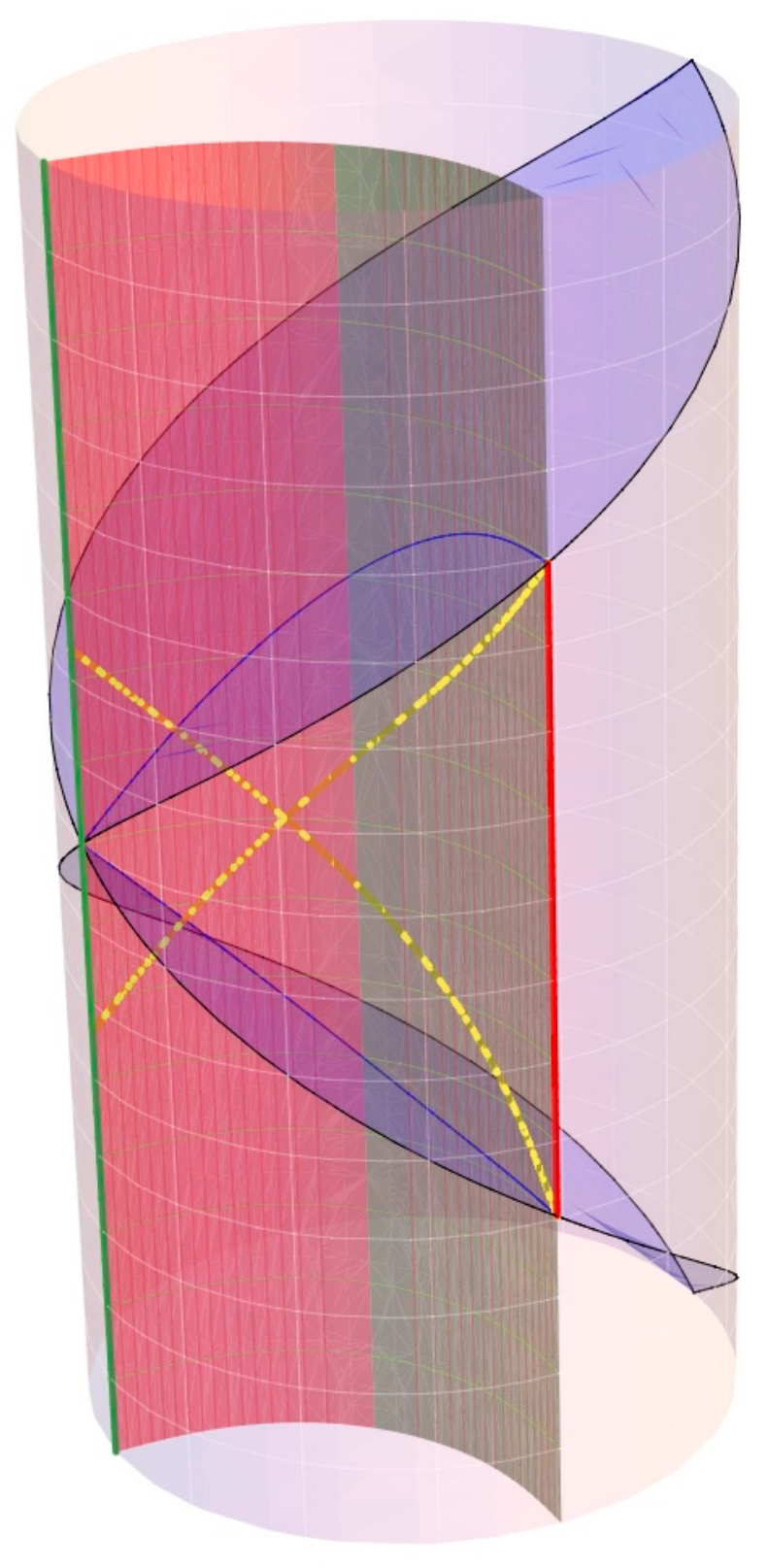}
\hspace{1cm}
\includegraphics[height=3in]{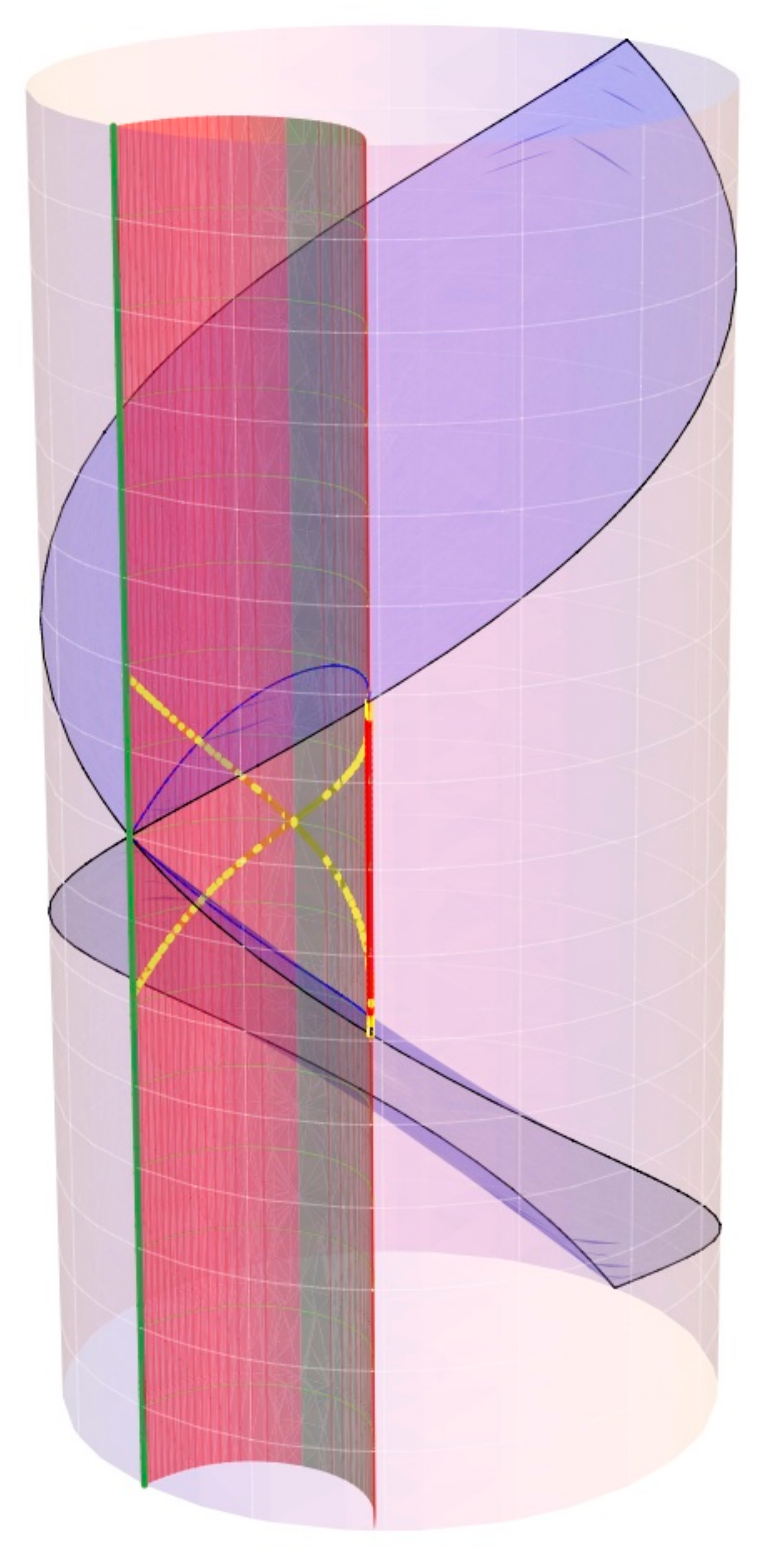}
\caption{
String worldsheet configuration when the endpoints are held fixed.  This corresponds to a single uniformly accelerating quark on Minkowski background with no \aq.  The blue surfaces correspond to the Poincare horizons, the string worldsheet is the vertical red and green surface (the two halves joined at $r=\rmin$ for ease of visualization), and  the red line at $\ph=\phq$ denotes the quark.  This is at $\phq = \pi/4$  (left),  $\phq = \pi/2$  (middle), and  $\phq = 3\pi/4$  (right).  The dark-blue curves are the intersection of the worldsheet with the Poincare horizon, which the dashed yellow curves (for greater ease of identification) denote the worldsheet event horizon.}
\label{fig:bentstring}
\end{center}
\end{figure}

The string profile, given by $\ph(r)$ for various values of $\phq$, is plotted on Poincare disk (depicting a  constant $\tg$ slice of global AdS) in the
left panel of \fig{fig:WSPoincDiskForced}, color-coded by acceleration (or equivalently $\phq$, or equivalently $\rmin$); the right panel shows the corresponding trajectory of the quark on the boundary Minkowski spacetime Penrose diagram.
The full string worldsheet in global AdS for three specific cases is given in \fig{fig:bentstring}. For ease of visualization we have flipped the plots in $\ph$ and indicated the left/right half of the worldsheet (given by \req{phiofrL} and \req{phiofrR}, respectively) by the green/red surfaces, which touch at $\rmin$.  The yellow and blue curves on the worldsheet correspond to event horizons and Poincare coordinate singularities, further discussed below.

%-----------------------
\paragraph{Worldsheet metric:}

The worldsheet metric $\gamma_{ab}$ is given by 
\begin{equation}
\dsWS^2 =  - (r^2 + 1) \,  d\tg^2 + \frac{1 + r^2 \, (r^2 + 1) \, \ph'(r)^2}{r^2+1} \, dr^2
\label{}
\end{equation}	
which, using \req{phiintgrnda}, can be re-expressed as
\begin{equation}
\dsWS^2 = 
- (r^2 + 1) \, d\tg^2
+ \frac{r^2}{r^2 \, (r^2 +1) - \rmin^2 \, (\rmin^2 +1)} \, dr^2 \ .
\label{bentWSmet}
\end{equation}	
Evaluating its curvature, we still have $R^{(\gamma)}_{ab} = -\gamma_{ab}$ as before, the Ricci scalar now has a mild dependence on $r$: %$R^{(\gamma)}$
\begin{equation}
R^{(\gamma)} = -2 \, \frac{(r^2 + 1)^2 + \rmin^2 \, (\rmin^2 +1)}{(r^2 + 1)^2}
\label{bentWSRicci}
\end{equation}	
which varies monotonically with $r$, attaining the (unit-radius) AdS value $-2$ as $r \to \infty$, and never exceeding twice this value around its deepest part.

% Figure 
%

%-----------------------
\paragraph{Worldsheet event horizon:}
In order to characterize the worldsheet event horizon, we first need to find  null curves (which are automatically null geodesics) in the worldsheet metric \req{bentWSmet}, and then find the specific set of null curves which end at the boundary points where the worldsheet intersects the Poincare horizon.  
We have to bear in mind that to cover the full region of interest, $\ph$ is double-valued in $r$.  Nevertheless, ignoring this subtlety for the moment, we can find null trajectories on the worldsheet by solving 
\begin{equation}
\frac{d\tg}{dr} = \frac{r}{\sqrt{r^2+1}} \, \frac{1}{\sqrt{r^2 \, (r^2 + 1)  - \rmin^2 \, (\rmin^2 + 1) }}
\label{}
\end{equation}	
which integrates to an expression in terms of elliptic functions:
\begin{equation}
\pm \tg(r) = \pi - \phq(\rmin) - \frac{1}{\sqrt{\rmin^2 + 1}}  F\left[ \sin^{-1}\sqrt{\frac{\rmin^2 + 1}{r^2 + 1}}\  , \  \frac{-\rmin^2}{\rmin^2 + 1}  \right]  \ .
\label{bentWShor}
\end{equation}	
Note that the time it takes a null trajectory on the worldsheet to get from $r=\rmin$ to $r = \infty$ is given by
\begin{equation}
\Delta \tg  = \frac{1}{\sqrt{\rmin^2 + 1}} \, K\left[ \frac{-\rmin^2}{\rmin^2 + 1} \right] 
\label{}
\end{equation}	
which has the expected behaviour of monotonically decreasing with $\rmin$, with  $\Delta \tg(\rmin=0) = \pi/2$ and $\Delta \tg(\rmin\to \infty) \to 0$.
Note that the horizon {\it always} reaches past $\rmin$.  In the limit of $\phq \to \pi$, the bifurcation point approaches $\rmin$, while in the opposite limit $\phq \to 0$, bifurcation point approaches $i^0$ of the Poincare patch.

%%%%%%%%%%%%%%%%%%%%%%%%%%%%%%%%%%
\bibliographystyle{JHEP}
%\bibliography{stringWS}
\providecommand{\href}[2]{#2}\begingroup\raggedright\endgroup

\end{document}